\pdfoutput=1
\documentclass[journal]{new-aiaa} 
\usepackage[utf8]{inputenc}

\usepackage{graphicx}
\usepackage{subcaption}
\usepackage{amsmath}
\usepackage[version=4]{mhchem}
\usepackage{siunitx}
\usepackage{longtable,tabularx, float}
\usepackage{braket}
\usepackage{makecell}
\usepackage{booktabs}
\usepackage{tikz}
\usetikzlibrary{arrows.meta, positioning, calc, shapes.geometric}
\usepackage{xcolor}
\usepackage[normalem]{ulem}

\title{Quantum-Inspired Evolutionary Optimization for Evaluation-Efficient Rotor Blade Aerodynamic Design}

\author{Jinwhuy Lee \footnote{Graduate Research Assistant, Department of Aerospace Engineering, jinwhuy@umd.edu}, Manmohan Thakur \footnote{Graduate Research Assistant, Department of Aerospace Engineering, mthakur@umd.edu} and James D. Baeder \footnote{Professor, Department of Aerospace Engineering, baeder@umd.edu}}
\affil{University of Maryland, College Park, Maryland 20740}
\author{Apurva Anand \footnote{Aerodynamic Engineer, anandaero747@gmail.com}}
\affil{Archer Aviation, San Jose, California, 95134}
\author{Kandula Eswara Sai Kumar \footnote{Product R\&D Manager, BQP, eswara.sai@bqpsim.com},  Rut Lineswala \footnote{CTO, rut.lineswala@bqpsim.com}, and Abhishek Chopra \footnote{CEO, abhishek.chopra@bqpsim.com}}
\affil{BosonQ Psi Corp., Syracuse, New York, 13202, USA}

\begin{document}

\maketitle

\begin{abstract}
This study investigates the application of quantum-inspired evolutionary optimization (QIO) to airfoil and rotor-blade aerodynamic design and compares its performance with that of a conventional genetic algorithm (GA) across different population sizes. QIO is a population-based evolutionary method that represents candidate solutions probabilistically, allowing the search distribution to explore multiple regions of the design space on classical computing hardware. To compare the two optimizers across problems with varying dimensionality and evaluation cost, two test cases are considered: a 20-variable airfoil shape optimization evaluated using a pretrained aerodynamic surrogate, and a 45-variable rotor blade optimization evaluated directly with UMARC-II. In both problems, QIO shows better convergence efficiency than GA as the population size decreases. For the airfoil optimization, QIO requires up to 49.6\% fewer evaluations to reach near-optimal solutions. For the small-population rotor blade optimization, QIO achieves an average figure of merit of 0.7411, compared with 0.7373 for GA. QIO also shows smaller standard deviations across independent runs, indicating more consistent performance. In addition, a neural-network-assisted approach is investigated for rotor blade optimization. Neural-network-informed initialization improves the final GA figure of merit from 0.7270 to 0.7290, although fresh-start QIO still achieves the highest value of 0.7409.
\end{abstract}

\section{Introduction}

Over the past decades, rotorcraft have attracted significant attention due to their unique vertical takeoff and landing capabilities and their expanding role in applications ranging from conventional helicopters to emerging urban air mobility (UAM) vehicles \cite{Filippone2021}. In rotorcraft performance, the rotor blade plays a central role because its aerodynamic characteristics directly govern the power required to sustain lift and propulsion under given flight conditions. Improved rotor blade design therefore remains a direct path toward better rotorcraft performance.

The design of a rotor blade involves the interaction of multiple geometric and aerodynamic parameters, including airfoil shape, twist distribution, and planform configuration. These parameters define a highly nonlinear, coupled, and multimodal design space, making aerodynamic optimization particularly challenging \cite{YPHong2024}. Furthermore, rotor blade analysis often requires computationally intensive simulations, especially when coupling aerodynamics with structural dynamics \cite{SJPark2025}. As a result, numerical optimization of rotor blades is both resource-intensive and computationally demanding, particularly when repeated evaluations are required \cite{Bailly2019}.

In general, two primary approaches have been adopted for numerical optimization. The first approach relies on gradient-based optimization methods, which are efficient in terms of the number of function evaluations required and well suited for problems with a large number of design variables \cite{Patel2022, Wang2019}. However, these methods require accurate gradient information and are inherently local, making them susceptible to converging to local minima in highly nonlinear design spaces \cite{Li2008}. The second approach relies on global optimization strategies to address the highly nonlinear and multimodal nature of rotor blade design problems. 
Among these, the Genetic Algorithm (GA) has been widely adopted for its robustness and ability to perform gradient-free global searches. GA is particularly suitable for rotor blade optimization problems, where the design space is complex and derivative information is either unavailable or unreliable \cite{Dawoon2022}. However, its exploration capability depends strongly on stochastic operators and population size. A large population helps maintain diversity but increases the number of aerodynamic evaluations per generation, whereas a small population reduces this cost but can converge prematurely.

In practical applications, to mitigate the computational cost, simplified aerodynamic models such as the lifting-line method \cite{Walsh1995, Joon2021} are often employed in conjunction with GA-based optimization. While this approach enables tractable optimization, the large number of evaluations required for convergence still limits its efficiency. Since the rotor blade performance evaluation becomes significantly more expensive with higher-fidelity models, surrogate-based approaches can be employed to improve efficiency \cite{Imiela2012, Cornelius2024}. In this framework, the complex design space is approximated using surrogate models, enabling rapid evaluation of candidate designs. Surrogate-based approaches reduce the cost of repeated evaluations, but their accuracy depends on the quality and coverage of the training data. In particular, surrogate models suffer from the curse of dimensionality, requiring a large number of training samples as the number of design variables increases \cite{Muneeb2024}. Furthermore, inaccurate predictions in undersampled regions may lead to erroneous identification of optimal solutions \cite{ZhongHua2012}. The construction of reliable surrogate models also incurs a high upfront computational cost, and their performance strongly depends on the choice of model type and tuning parameters \cite{Wilke2019}.

Previous studies have shown that airfoil geometry can strongly affect rotor blade performance \cite{Muneeb2026, Apurva2026}. However, it also greatly increases the dimensionality of the design space. Parameterizing airfoil geometry typically requires many design variables, exacerbating the curse of dimensionality and making global optimization substantially more challenging. To address this issue, previous studies have proposed strategies to reduce the complexity of the optimization problem. One common approach is successive optimization, in which airfoil shapes are optimized separately before optimizing blade-level parameters such as twist and planform geometry \cite{Gunther2023}. Another approach involves airfoil selection, in which the design problem is simplified by selecting from a predefined database of candidate airfoils rather than directly optimizing the airfoil shape \cite{Apurva2026VFS, Muneeb2025ERF}. While these methods effectively reduce the dimensionality of the optimization problem, they inherently restrict the exploration of the full design space and may prevent the identification of globally optimal solutions.

To retain the flexibility of a fully coupled design space while limiting reliance on surrogate models and their associated prediction uncertainty and training data requirements, the present study considers quantum-inspired optimization (QIO) as an alternative strategy to improve the efficiency of global search. Quantum-inspired evolutionary algorithms introduced by Han and Kim employ qubit-based probabilistic representations rather than explicitly encoding candidate solutions in the search population \cite{han2002quantum}. This representation retains information about multiple possible states for each design variable, thereby supporting broader exploration of the design space while running entirely on classical hardware. Building on this concept, Kumar et al. reported that QIO reached target fitness values with fewer function evaluations and exhibited more consistent convergence than GA for the Ackley, Rosenbrock, and Rastrigin benchmark functions \cite{kumar2024benchmarking}. Motivated by these findings, the present study addresses three central research questions:
\begin{itemize}
    \item {Can QIO maintain effective global search in high-dimensional aerodynamic design when the population size is strictly constrained?}
    \item {How does population reduction impact the run-to-run consistency and resulting blade geometry of QIO compared to conventional GA?}
    \item {Does fresh-start QIO eliminate the need for offline surrogate training when compared against surrogate-informed hybrid evolutionary frameworks?}
\end{itemize}

\section{Methodology}
\begin{figure}[hbt]
    \centering
    \includegraphics[width=1.00\textwidth]{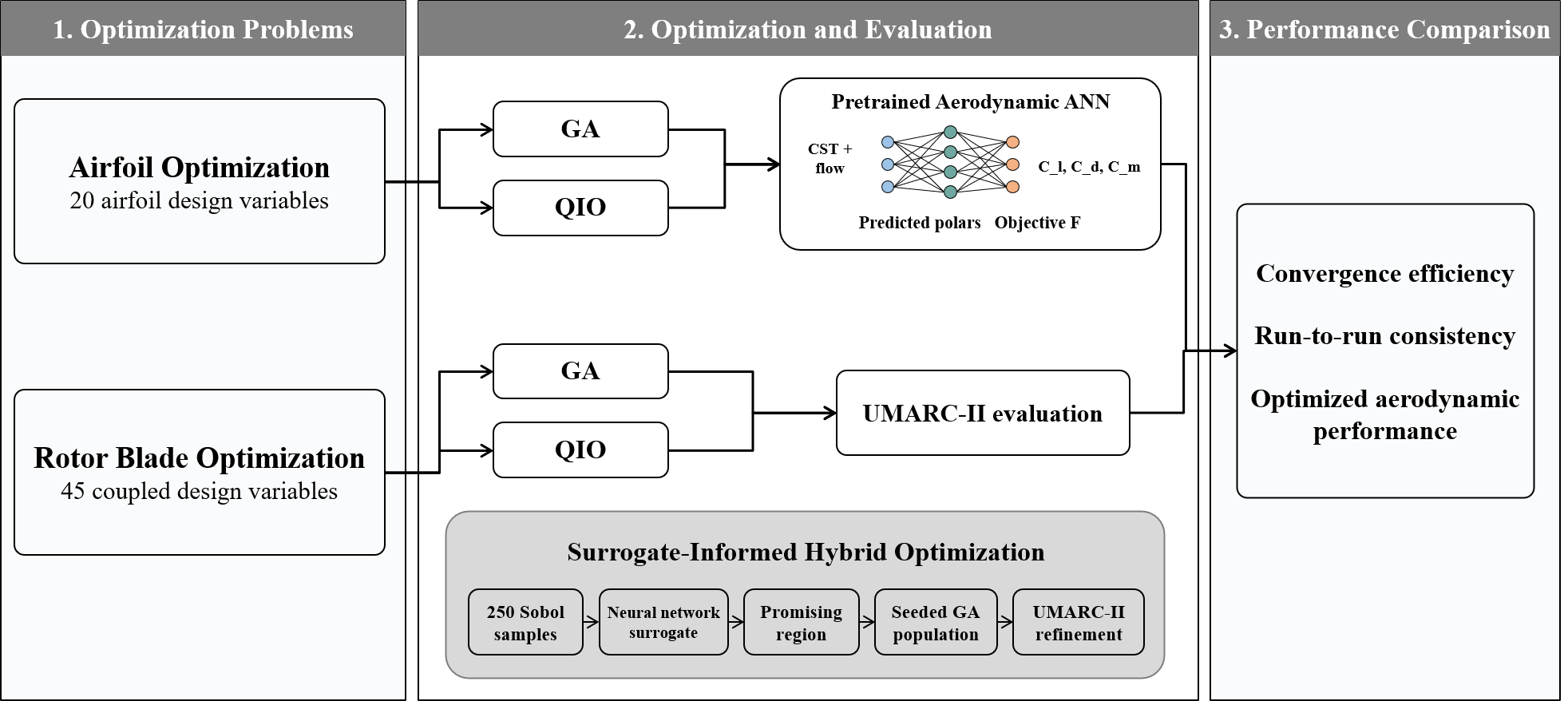}
    \caption{Overview of the methodology for the airfoil and rotor blade optimization problems}
    \label{fig:methodology_framework}
\end{figure}

Figure \ref{fig:methodology_framework} summarizes the methodology used for the two aerodynamic optimization problems considered in this study. For both the airfoil and rotor blade problems, GA and QIO are evaluated, while a surrogate-informed hybrid strategy is additionally examined for the rotor-blade problem by using a neural-network-predicted promising region to initialize the population before UMARC-II refinement.

\subsection{Optimization Problem Definitions}
\subsubsection{Airfoil Optimization}
The first design problem considered in this study is an airfoil aerodynamic optimization problem. This problem is used to evaluate the optimizer's performance in a high-dimensional airfoil-shape design space while keeping the aerodynamic evaluation computationally efficient via a surrogate model.

The baseline airfoil is the VR-8 airfoil. The airfoil geometry is parameterized using a Chebyshev-based class-shape transformation (CST) representation \cite{Apurva2024}. The airfoil surface is expressed as
\begin{equation}
    y(x) = C(x) S(x),
\end{equation}
where $C(x)$ is the class function and $S(x)$ is the shape function. The class function is defined as
\begin{equation}
    C(x) = \sqrt{x}(1-x),
\end{equation}
and the shape function is represented using Chebyshev polynomials as
\begin{equation}
    S(x) = \sum_{i=0}^{n} c_i T_i(x),
\end{equation}
where $T_i(x)$ is the $i$th Chebyshev polynomial and $c_i$ is the corresponding shape coefficient. In this work, 10 design variables are used for the upper surface and 10 for the lower surface, resulting in a total of 20 airfoil design variables. The aerodynamic coefficients of each candidate airfoil were evaluated using a pretrained feed-forward neural-network surrogate developed from a high-fidelity Reynolds-averaged Navier-Stokes (RANS) CFD database. A vectorized network architecture is employed to predict the complete aerodynamic polars rather than individual operating points, and proper orthogonal decomposition is used to reduce the dimensionality of the output data before network training. The aerodynamic coefficients corresponding to the specified design condition are extracted from the predicted polars for objective evaluation. Further details regarding the CFD database, network architecture, data preprocessing, training procedure, and validation are provided in Ref. \cite{Apurva2026}.

{The airfoil optimization was conducted at $M=0.6$ and $\alpha=1^\circ$, which were selected as representative operating conditions for the outboard region of a rotor blade.} The airfoil objective function is formulated as a combined objective that rewards a high lift-to-drag ratio and lift coefficient, while penalizing drag and excessive pitching moment. The scalar objective is defined as
\begin{equation}
    F =
    -w_{L/D}\times\frac{L}{D}
    -w_{C_l}\times C_l
    +w_{C_d}\times C_d
    +w_{C_m}\times \left[\max(0, C_m-C_{m_0})\right]^2,
    \label{eq:airfoil_objective}
\end{equation}
where $w_{L/D}=1.0$, $w_{C_l}=0.2$, $w_{C_d}=0.5$, and $w_{C_m}=50$ are the weighting coefficients. The optimization is subject to a lift constraint,
\begin{equation}
    C_l \geq C_{l,\mathrm{baseline}} - 0.1,
\end{equation}
where $C_{l,\mathrm{baseline}}$ is the lift coefficient of the baseline VR-8 airfoil at the same flow condition. In addition, the optimized airfoils are thickness-scaled to match the baseline airfoil thickness, ensuring that the resulting geometries remain practically comparable to the original airfoil.

{For the airfoil optimization, the population sizes were selected to represent three levels of population reduction: a large-population reference with 80 candidates, an intermediate case with 40 candidates, and a strongly reduced case with 10 candidates. Rather than tuning the computational budget independently for each case, the total number of surrogate evaluations was fixed at 16,000 to isolate the effect of population size on optimizer performance.} The population size and number of generations for each case are provided in Table \ref{tab:airfoil_settings}. Each case is repeated using 15 different random seeds, and GA and QIO use the same initial population for each corresponding seed.

\begin{table}[H]
    \centering
    \caption{Optimization settings for the airfoil design problem}
    \label{tab:airfoil_settings}
    \begin{tabular}{ccccc}
        \hline
        Case & Population size & Generations & Total evaluations & Number of seeds \\
        \hline
        Large population & 80 & 200 & 16,000 & 15 \\
        Moderate population & 40 & 400 & 16,000 & 15 \\
        Small population & 10 & 1,600 & 16,000 & 15 \\
        \hline
    \end{tabular}
\end{table}

\subsubsection{Rotor-Blade Optimization}
\begin{figure}[H]
    \centering
    \includegraphics[width=0.7\linewidth]{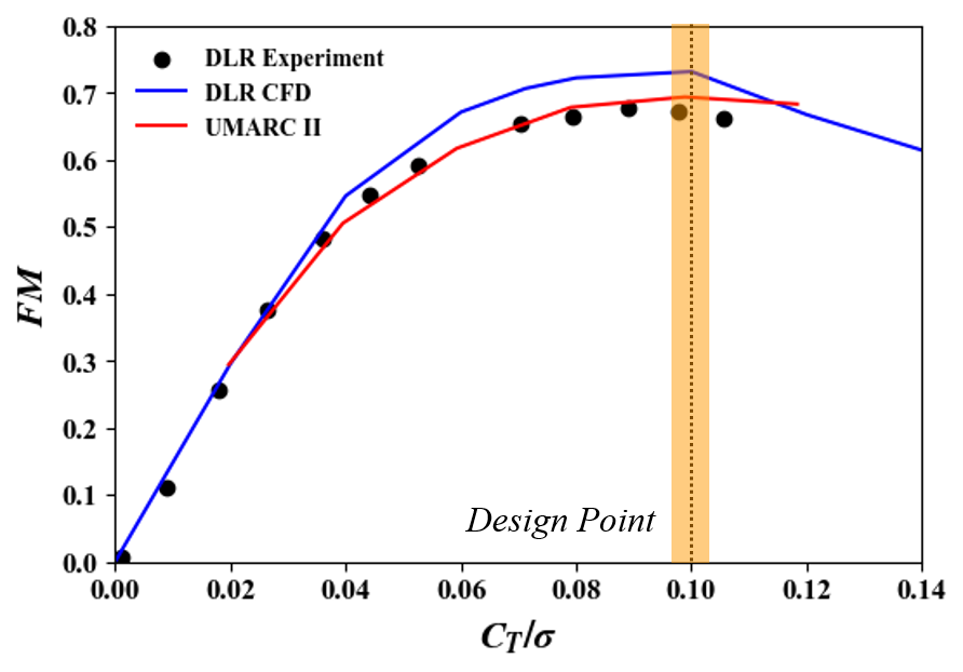}
    \caption{
    Validation of the UMARC-II hover performance for the baseline HART-II rotor
    }
    \label{fig:hartii_umarc}
\end{figure}

{The second design problem considers the aerodynamic optimization of the HART-II model-scale rotor, which is based on the hingeless Bo-105 rotor and employs a rectangular baseline planform \cite{vanderWall2003}. Hover performance is evaluated using UMARC-II, a comprehensive rotorcraft analysis solver developed at the University of Maryland \cite{Sebastian2010,Muneeb2025Scitech}, and is quantified by the figure of merit (FM). The figure of merit is defined as
\begin{equation}
    Figure\ of\ Merit\ (FM) = \frac{C_T^{3/2}}{\sqrt{2}C_P}
\end{equation}
where $C_T$ and $C_P$ are the thrust and power coefficients. Among the aerodynamic modeling options available in UMARC-II, the linear inflow model was selected to make repeated design evaluations across multiple population sizes and random seeds computationally tractable.
Before performing the optimization, the UMARC-II model was evaluated to determine whether it could reproduce the hover-performance trend of the baseline HART-II rotor. In Fig. ~\ref{fig:hartii_umarc}, the UMARC-II prediction is compared with the experimental measurements and CFD results reported in Ref.~\cite{JOD2022} over a range of hover-loading conditions.} As shown in Fig.~\ref{fig:hartii_umarc}, UMARC-II captures the overall variation of FM with $C_T/\sigma$ and provides reasonable agreement with the reference data near the optimization condition of $C_T/\sigma=0.10$.

\begin{figure}[H]
    \centering
    \includegraphics[width=0.8\linewidth]{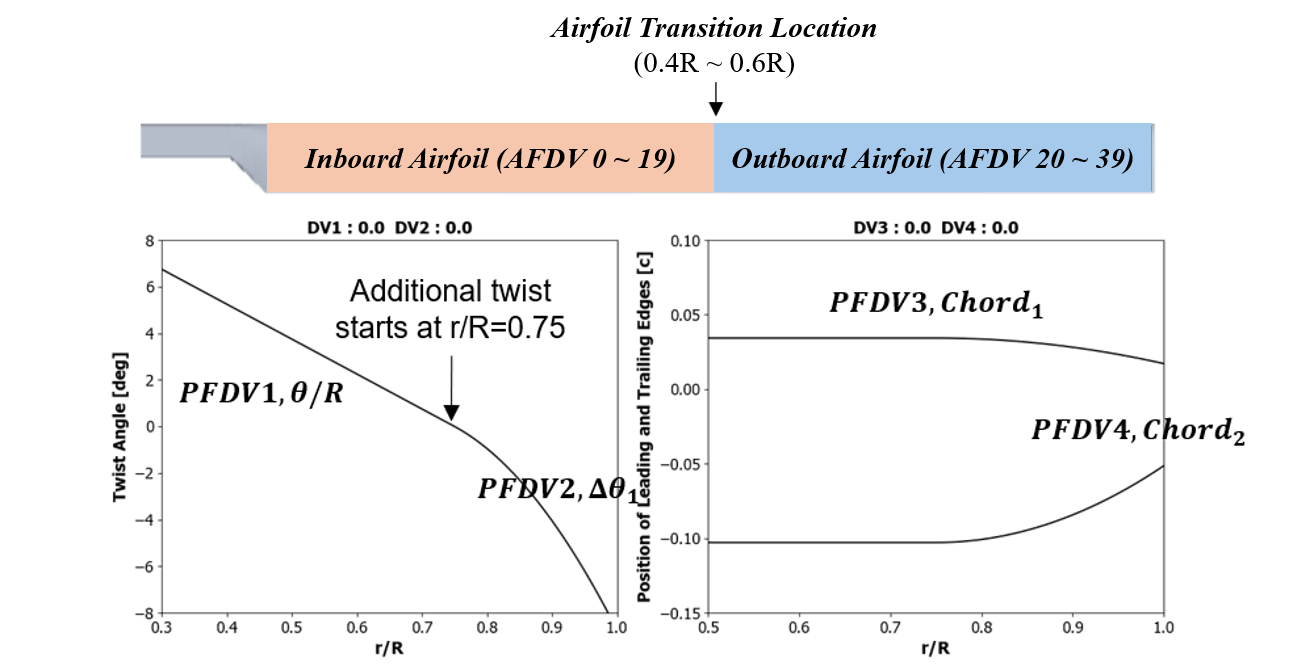}
    \caption{Rotor blade design variables, including inboard and outboard airfoil definitions, transition location, twist, and chord distribution}
    \label{fig:rotor_design_variables}
\end{figure}

Figure~\ref{fig:rotor_design_variables} summarizes how design variables are defined to construct the rotor-blade geometry.
The rotor blade design variables consist of airfoil shape variables (AFDV), an airfoil transition location, and planform design variables (PFDV). Two airfoils are defined along the blade: an inboard airfoil and an outboard airfoil. The baseline airfoil coefficient vectors for both sections are obtained by fitting the Chebyshev-based CST representation to the NACA 23012 airfoil. The transition location between the two airfoils is also treated as a design variable and is constrained from 0.4$R$ to 0.6$R$, where $R$ is the rotor radius.

In addition to the airfoil variables, four planform design variables are included. These variables modify the blade's twist and chord distributions. The additional twist is applied starting from $r/R=0.75$, and the chord distribution is modified using two chord-related design variables\cite{JOD2023}. {Smooth and continuous spanwise distributions are reconstructed from these control points using cubic-spline interpolation. After the chord distribution is reconstructed, the chord lengths are rescaled to preserve the thrust-weighted solidity of the baseline blade.} Since each airfoil has 20 design variables, the two airfoils contribute 40 design variables. Including one transition-location variable and four planform variables gives a total of 45 design variables.

The objective of the rotor blade optimization is to maximize the hover figure of merit. For the rotor blade optimization, two population-size cases are considered. The population size and number of generations for each case are provided in Table \ref{tab:rotor_settings}. Each case is repeated using 10 different random seeds, and the same optimization settings are used for GA and QIO to ensure a consistent comparison between the two optimizers.

\begin{table}[H]
    \centering
    \caption{Optimization settings for the rotor blade design problem}
    \label{tab:rotor_settings}
    \begin{tabular}{cccc}
        \hline
        Case & Population size & Generations & Number of seeds \\
        \hline
        Small population & 30 & 50 & 10 \\
        Large population & 256 & 50 & 10 \\
        \hline
    \end{tabular}
\end{table}

\subsection{Choice of Optimization Solvers}

\subsubsection{Quantum-Inspired Evolutionary Optimization (QIO)}

\begin{figure}[htb]
\centering
\begin{tikzpicture}[
    node distance=0.65cm and 1.0cm,
    font=\small\sffamily,
    block/.style={
        draw=black,
        thick,
        rounded corners=4pt,
        align=center,
        inner sep=6pt,
        minimum width=5.6cm,
        minimum height=0.9cm,
        line width=0.8pt
    },
    arrow/.style={
        -{Stealth[length=2.5mm, width=1.8mm]},
        thick,
        line width=0.9pt
    }
]

    \node (init)    [block] {Population Initialization via \\ Hadamard Gate};
    \node (circuit) [block, below=of init]    {Creation of Parametric Quantum Circuit \\ with Rotation Gate};
    \node (meas)    [block, below=of circuit]   {Measurement to get Population};
    \node (fit)     [block, below=of meas]    {Fitness Calculation};
    \node (best)    [block, below=of fit]     {Find the Best Population};
    \node (update)  [block, below=of best]    {Update the Parametric Gates for Child\\Population};

    \draw [arrow] (init) -- (circuit);
    \draw [arrow] (circuit) -- (meas);
    \draw [arrow] (meas) -- (fit);
    \draw [arrow] (fit) -- (best);
    \draw [arrow] (best) -- (update);
    \draw [arrow] (update.west) -- ++(-0.8,0) |- (meas.west);

\end{tikzpicture}
\caption{Flow chart of QIO algorithm}
\label{fig:QIO_flowchart}
\end{figure}
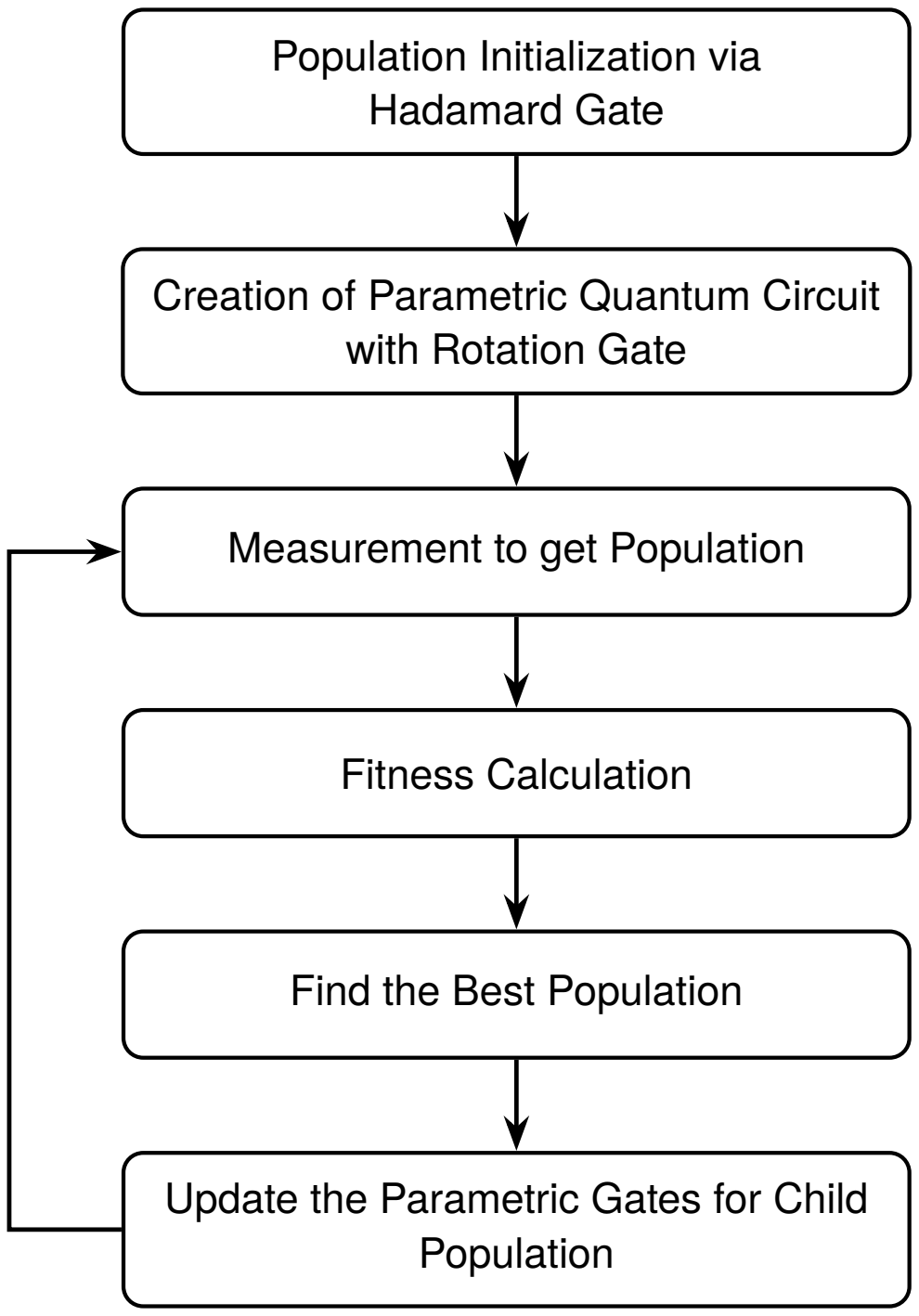

Quantum-inspired evolutionary optimization (QIO) is an evolutionary optimization approach that adopts quantum-mechanical concepts, including qubits, superposition, quantum gates, and measurement, while running on classical computing hardware. QIO is employed as a stochastic, population-based optimizer and benchmarked against a conventional genetic algorithm (GA) for the aerodynamic design problems considered in this study. The QIO workflow adopted herein is summarized in Fig.~\ref{fig:QIO_flowchart}.

In a classical genetic algorithm, each chromosome represents one deterministic candidate solution in the design space. In contrast, QIO represents each design variable as a probabilistic quantum state. A quantum gene can be expressed as
\begin{equation}
    \ket{\psi} = \alpha \ket{0} + \beta \ket{1},
\end{equation}
where $\alpha$ and $\beta$ are probability amplitudes associated with the basis states $\ket{0}$ and $\ket{1}$, respectively. The probabilities of observing the corresponding classical states are given by $|\alpha|^2$ and $|\beta|^2$, with
\begin{equation}
    |\alpha|^2 + |\beta|^2 = 1.
\end{equation}
For an individual chromosome represented by $m$ qubits, the quantum chromosome can be written as
\begin{equation}
    q_j =
    \begin{bmatrix}
    \alpha_1 & \alpha_2 & \cdots & \alpha_m \\
    \beta_1  & \beta_2  & \cdots & \beta_m
    \end{bmatrix}.
\end{equation}
Thus, a quantum chromosome encodes a probability distribution over possible classical states rather than one fixed candidate. This probabilistic representation allows a single chromosome to retain information across multiple candidate solutions, whereas a classical chromosome corresponds to a single fixed state.

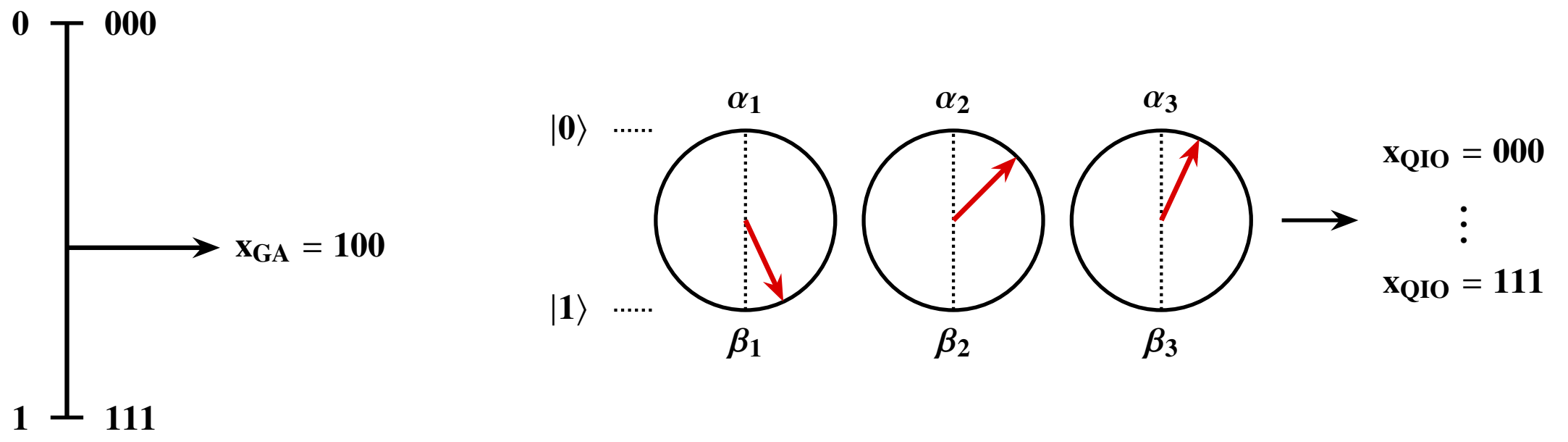
\begin{figure}[htbp]
\centering
\begin{tikzpicture}[
    font=\small,
    >=Stealth
]

    \begin{scope}[shift={(0,0)}]
        \draw[line width=1.2pt] (0, 1.8) -- (0, -1.8);
        
        \draw[line width=1.0pt] (-0.15, 1.8) -- (0.15, 1.8);
        \node[left=2pt] at (-0.15, 1.8) {\textbf{0}};
        \node[right=2pt] at (0.15, 1.8) {\textbf{000}};
        
        \draw[line width=1.0pt] (-0.15, -1.8) -- (0.15, -1.8);
        \node[left=2pt] at (-0.15, -1.8) {\textbf{1}};
        \node[right=2pt] at (0.15, -1.8) {\textbf{111}};
        
        \coordinate (ga_pt) at (0, -0.25);
        \draw[line width=1.2pt, -{Stealth[length=2.5mm]}] (ga_pt) -- ++(1.4, 0) 
            node[right] {$\mathbf{x}_{\mathbf{GA}} = \mathbf{100}$};
        
        \node at (0.6, -2.6) {(a) Genetic Algorithm};
    \end{scope}

    \begin{scope}[shift={(6.2,0)}]
        \def\R{0.82} 
        
        \node[left] at (-1.3, \R) {$|\mathbf{0}\rangle$};
        \draw[densely dotted, thick] (-1.2, \R) -- (-0.85, \R);
        
        \node[left] at (-1.3, -\R) {$|\mathbf{1}\rangle$};
        \draw[densely dotted, thick] (-1.2, -\R) -- (-0.85, -\R);

        \coordinate (c1) at (0, 0);
        \draw[line width=1.0pt] (c1) circle (\R);
        \draw[densely dotted, thick] ($(c1)+(0,-\R)$) -- ($(c1)+(0,\R)$);
        \node[above=1pt] at ($(c1)+(0,\R)$) {$\boldsymbol{\alpha_1}$};
        \node[below=1pt] at ($(c1)+(0,-\R)$) {$\boldsymbol{\beta_1}$};
        \draw[-{Stealth[length=2.4mm]}, line width=1.3pt, red!85!black] (c1) -- ++(-65:\R);

        \coordinate (c2) at (1.9, 0);
        \draw[line width=1.0pt] (c2) circle (\R);
        \draw[densely dotted, thick] ($(c2)+(0,-\R)$) -- ($(c2)+(0,\R)$);
        \node[above=1pt] at ($(c2)+(0,\R)$) {$\boldsymbol{\alpha_2}$};
        \node[below=1pt] at ($(c2)+(0,-\R)$) {$\boldsymbol{\beta_2}$};
        \draw[-{Stealth[length=2.4mm]}, line width=1.3pt, red!85!black] (c2) -- ++(45:\R);

        \coordinate (c3) at (3.8, 0);
        \draw[line width=1.0pt] (c3) circle (\R);
        \draw[densely dotted, thick] ($(c3)+(0,-\R)$) -- ($(c3)+(0,\R)$);
        \node[above=1pt] at ($(c3)+(0,\R)$) {$\boldsymbol{\alpha_3}$};
        \node[below=1pt] at ($(c3)+(0,-\R)$) {$\boldsymbol{\beta_3}$};
        \draw[-{Stealth[length=2.4mm]}, line width=1.3pt, red!85!black] (c3) -- ++(65:\R);

        \draw[-{Stealth[length=2.5mm]}, line width=1.0pt] (4.9, 0) -- (5.6, 0);

        \node[right, align=center] at (5.7, 0) {
            $\mathbf{x}_{\mathbf{QIO}} = \mathbf{000}$\\[2pt]
            $\boldsymbol{\vdots}$\\[2pt]
            $\mathbf{x}_{\mathbf{QIO}} = \mathbf{111}$
        };

        \node at (2.4, -2.6) {(b) Quantum-Inspired Optimization};
    \end{scope}

\end{tikzpicture}
\caption{Deterministic chromosome representation in (a) GA and probabilistic chromosome representation in (b) QIO}
\label{fig:chromosome_representation}
\end{figure}

The difference between the GA and QIO of chromosome representations is illustrated
in Fig.~\ref{fig:chromosome_representation} using a three-gene example. In the conventional GA representation shown in Fig.~\ref{fig:chromosome_representation} (a), the chromosome $\boldsymbol{x}_{GA}=100$ corresponds to a single deterministic candidate solution. In contrast, each gene in the QIO chromosome shown in Fig.~\ref{fig:chromosome_representation} (b) is described by a pair of probability amplitudes, $(\alpha_i,\beta_i)$, which define the probabilities of observing the corresponding binary states.

At the beginning of the optimization, the quantum population is initialized to provide an unbiased sampling of the design space. A measurement operation is then performed to collapse each quantum chromosome into a classical chromosome. The resulting classical population is mapped to the physical design variable space and evaluated using the corresponding objective function. After obtaining the fitness values, the best individual is identified and used to update the quantum population. The quantum state is updated using a rotation gate. In this study, the rotation operation is expressed as
\begin{equation}
    \begin{bmatrix}
    \alpha^t \\
    \beta^t
    \end{bmatrix}
    =
    \begin{bmatrix}
    \cos{\theta} & -\sin{\theta} \\
    \sin{\theta} & \cos{\theta}
    \end{bmatrix}
    \begin{bmatrix}
    \alpha^{t-1} \\
    \beta^{t-1}
    \end{bmatrix},
\end{equation}
where $\theta$ is the rotation angle, and the superscript $t$ denotes the generation index. The rotation angle is determined by comparing each bit of the measured individual with the corresponding bit of the best solution. If the measured bit differs from the best solution, the rotation direction is chosen such that the probability of generating a bit similar to the best solution is increased in the following generation. If the bit is already consistent with the best solution, no rotation is applied.

In the QIO solver configuration, the magnitude of the rotation update is controlled by a user-specified rotation-step parameter, denoted here as $\Delta\theta$. A larger value produces a stronger update and favors faster exploitation, whereas a smaller value provides a more gradual update and retains greater exploration. The airfoil and rotor-blade optimization cases used constant rotation-step parameters of $\Delta\theta=0.05\pi$ ($0.1571$ rad) and $\Delta\theta=0.10\pi$ ($0.3142$ rad), respectively.

This update strategy partially aligns the quantum states toward the best solution while retaining nonzero probabilities of generating alternative states. Therefore, QIO can maintain a balance between exploiting promising regions and exploring other regions in the design space. This feature is particularly attractive for optimization problems where the available population size is limited.

\subsubsection{Genetic Algorithm}
A conventional genetic algorithm is used as the baseline optimizer for comparison. GA is a population-based evolutionary algorithm in which each chromosome directly represents a candidate solution in the design space. The algorithm begins with an initial population, evaluates each individual's fitness, and evolves the population through selection, crossover, mutation, and survival operations.

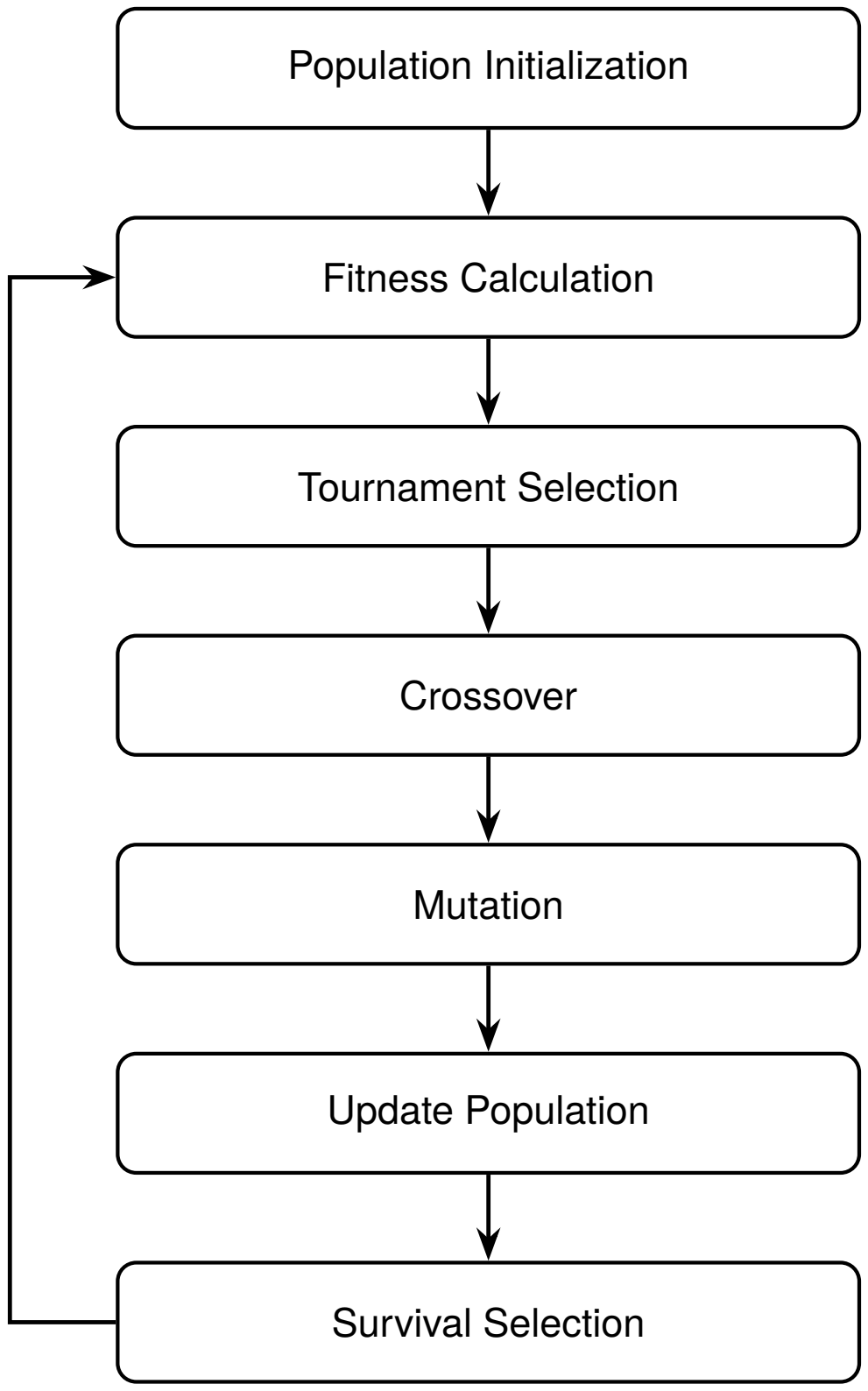
\begin{figure}[htbp]
\centering
\begin{tikzpicture}[
    node distance=0.65cm and 1.0cm,
    font=\small\sffamily,
    block/.style={
        draw=black,
        thick,
        rounded corners=4pt,
        align=center,
        inner sep=6pt,
        minimum width=5.6cm,
        minimum height=0.9cm,
        line width=0.8pt
    },
    arrow/.style={
        -{Stealth[length=2.5mm, width=1.8mm]},
        thick,
        line width=0.9pt
    }
]

    \node (init_pop)       [block] {Population Initialization};
    \node (fit1)       [block, below=of init_pop] {Fitness Calculation};
    \node (T_sel)  [block, below=of fit1]       {Tournament Selection};
    \node (cross)  [block, below=of T_sel]       {Crossover};
    \node (mut)       [block, below=of cross]  {Mutation};
    \node (update_pop) [block, below=of mut]       {Update Population};
    \node (sur_sel) [block, below=of update_pop]       {Survival Selection};

    \draw [arrow] (init_pop) -- (fit1);
    \draw [arrow] (fit1) -- (T_sel);
    \draw [arrow] (T_sel) -- (cross);
    \draw [arrow] (cross) -- (mut);
    \draw [arrow] (mut) -- (update_pop);
    \draw [arrow] (update_pop) -- (sur_sel);

    \draw [arrow] (sur_sel.west) -- ++(-0.8,0) |- (fit1.west);

\end{tikzpicture}
\caption{Flow chart of GA algorithm}
\label{fig:GA}
\end{figure}

In the present study, the GA follows a generational evolutionary procedure as shown in Fig.~\ref{fig:GA}. At each generation, parent individuals are selected from the current population using a fitness-based selection strategy. Crossover is then applied to generate offspring by recombining information from selected parents. Mutation is subsequently applied to introduce random perturbations into the offspring population, thereby maintaining diversity. The new candidate solutions are evaluated, and the next population is formed by retaining the fittest individuals.

Unlike QIO, GA's exploration capability is primarily governed by crossover and mutation probabilities. Therefore, its performance can be sensitive to operator settings and population size. In particular, when the population size is small, GA may lose diversity more rapidly and converge prematurely to suboptimal design regions. This behavior motivates the comparison with QIO under reduced-population optimization settings. In this study, the airfoil GA was implemented using the pymoo library \cite{blank2020pymoo}, whereas the rotor-blade GA was implemented using DEAP \cite{fortin2012deap}. The GA parameters were selected based on commonly used settings for continuous design optimization and are summarized in Table \ref{tab:ga_settings}. The parameters were kept fixed across all population-size cases to isolate the effects of the optimizer and population size.

\begin{table}[H]
    \centering
    \caption{Genetic algorithm implementations and operator settings}
    \label{tab:ga_settings}
    \small
    \renewcommand{\arraystretch}{1.25}
    \setlength{\tabcolsep}{8pt}

    \begin{tabular}{@{}ccc@{}}
        \toprule
        \text{Design problem} &
        \text{Crossover} &
        \text{Mutation} \\
        \midrule

        Airfoil optimization &
        \makecell[c]{
            Simulated binary crossover (SBX) \\
            $p_c=0.90,\ \eta_c=15$
        } &
        \makecell[c]{
            Gaussian mutation \\
            $\sigma=0.02$
        } \\
        \addlinespace[0.6em]

        Rotor-blade optimization &
        \makecell[c]{
            Bounded simulated binary crossover \\
            $p_c=0.90,\ \eta_c=15$
        } &
        \makecell[c]{
            Polynomial mutation \\
            $p_m=0.20,\ \eta_m=15$,\ $p_{m,\mathrm{var}}=1/45$
        } \\

        \bottomrule
    \end{tabular}
\end{table}

\subsubsection{Surrogate-Informed Hybrid Optimization}
Neural-network-based surrogate models provide an efficient means of exploring high-dimensional rotor design spaces because they can evaluate candidate designs at a substantially lower cost than direct physics-based analysis. However, constructing a reliable surrogate requires an initial dataset generated using the original analysis model. Furthermore, since the surrogate provides predicted rather than directly computed objective values, its predictions carry uncertainty, particularly in regions insufficiently represented in the training data. 

To address these limitations, previous studies have combined neural-network surrogate models with evolutionary algorithms to accelerate aerodynamic design optimization while reducing the number of expensive high-fidelity evaluations \cite{Hacioglu2007, rai2002hybrid, xu2019optimization}. Inspired by these approaches, the present study employs the surrogate-informed hybrid framework. First, a neural-network surrogate trained using a relatively small high-fidelity dataset is used to screen the design space and identify a promising region. Candidate designs sampled from this region are then used to seed the initial population, thereby accelerating the early global-search stage. The subsequent evolutionary refinement is performed using UMARC-II evaluations rather than surrogate predictions. 

\begin{figure}[H]
    \centering
    \includegraphics[width=1.0\linewidth]{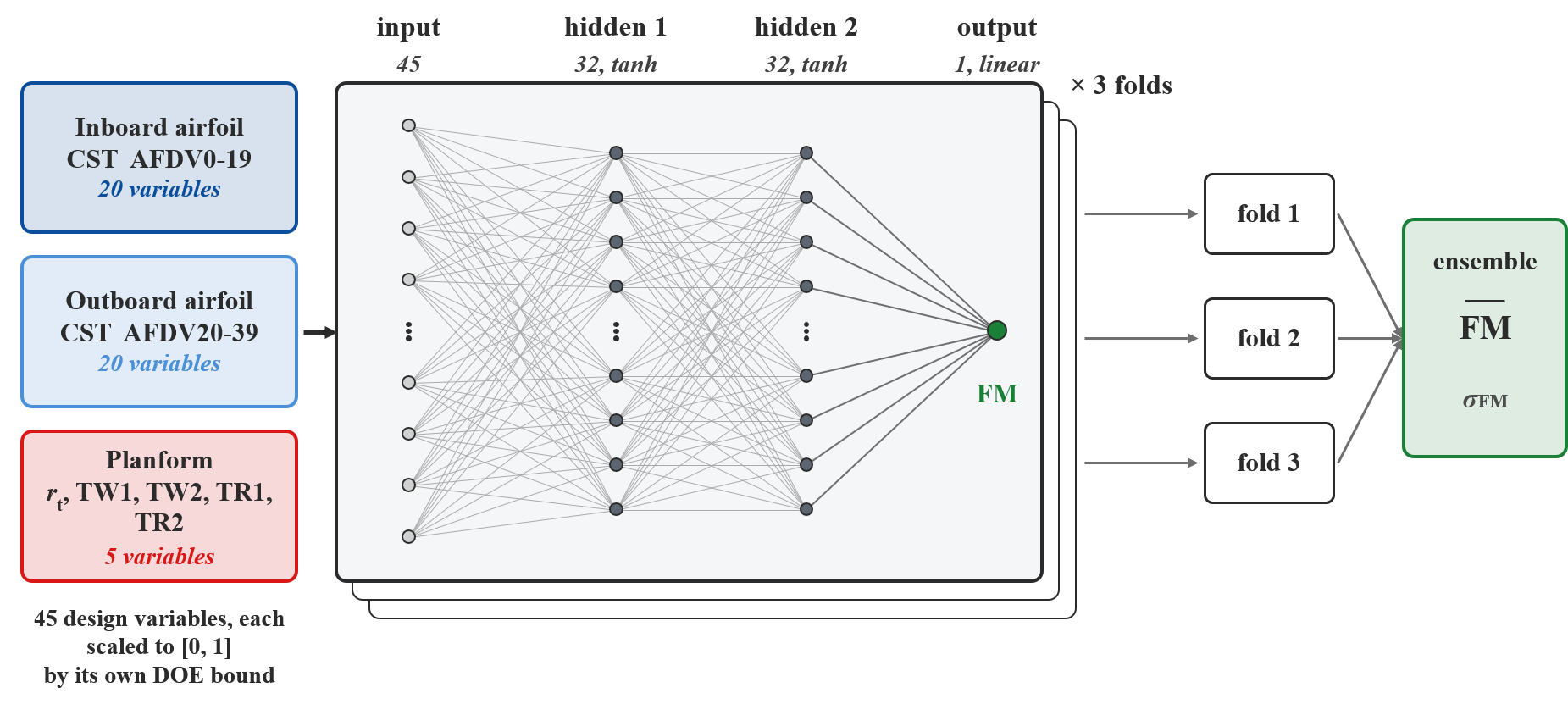}
    \caption{Architecture of the neural network}
    \label{fig:nn_architecture}
\end{figure}

To generate a neural network model, a total of 250 rotor designs were generated using Sobol sampling \cite{sobol1967distribution, matousek1998discrepancy} over the normalized 45-dimensional design space. Each design was evaluated using UMARC-II, and the corresponding hover figure of merit was used as the target output for neural-network training. The surrogate model is constructed as a fully connected multilayer perceptron with 45 input variables and a single output representing the hover figure of merit. The network contains two hidden layers with 32 neurons each and uses the hyperbolic tangent activation function. The network weights are determined by minimizing the mean squared error between the UMARC-II results and the neural network predictions. The overall architecture of the neural network is shown in Fig.~\ref{fig:nn_architecture}.

\begin{figure}[hbt]
    \centering
    \includegraphics[width=0.6\linewidth]{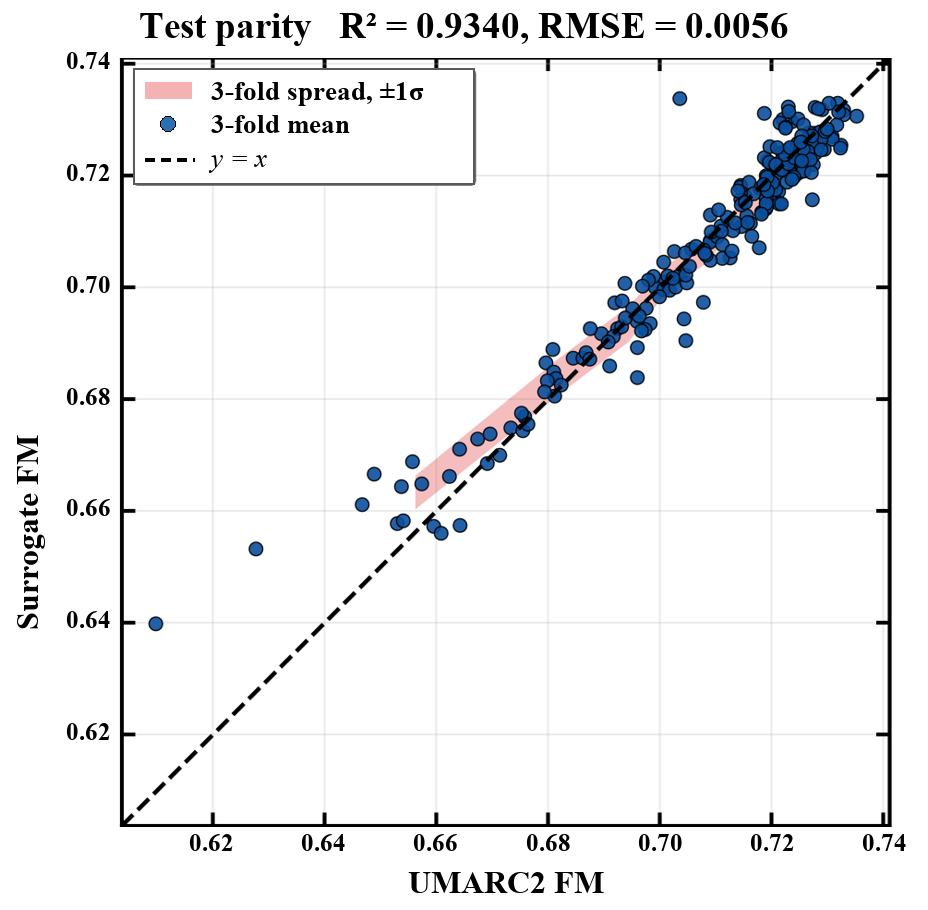}
    \caption{Parity plot of the surrogate model}
    \label{fig:nn_parity}
\end{figure}

Three-fold cross-validation was used to evaluate the model. Three separate networks were trained, and their predictions were averaged to obtain the final surrogate prediction. The variation among the three predictions was used to quantify the model spread. The resulting surrogate achieved a test coefficient of determination of $R^2 = 0.9340$ and a root-mean-square error of $RMSE_{FM} = 0.0056$. Figure~\ref{fig:nn_parity} compares the surrogate predictions with the corresponding UMARC-II results. Most predictions remain close to the ideal $y=x$ line, indicating that the model captures the overall relationship between the 45 rotor design variables and hover figure of merit. The spread among the cross-validation models also illustrates the uncertainty associated with the surrogate predictions.
\section{Results}
\subsection{Airfoil Optimization Results}
\begin{figure}[hbt]
    \centering
    \includegraphics[width=0.85\linewidth]{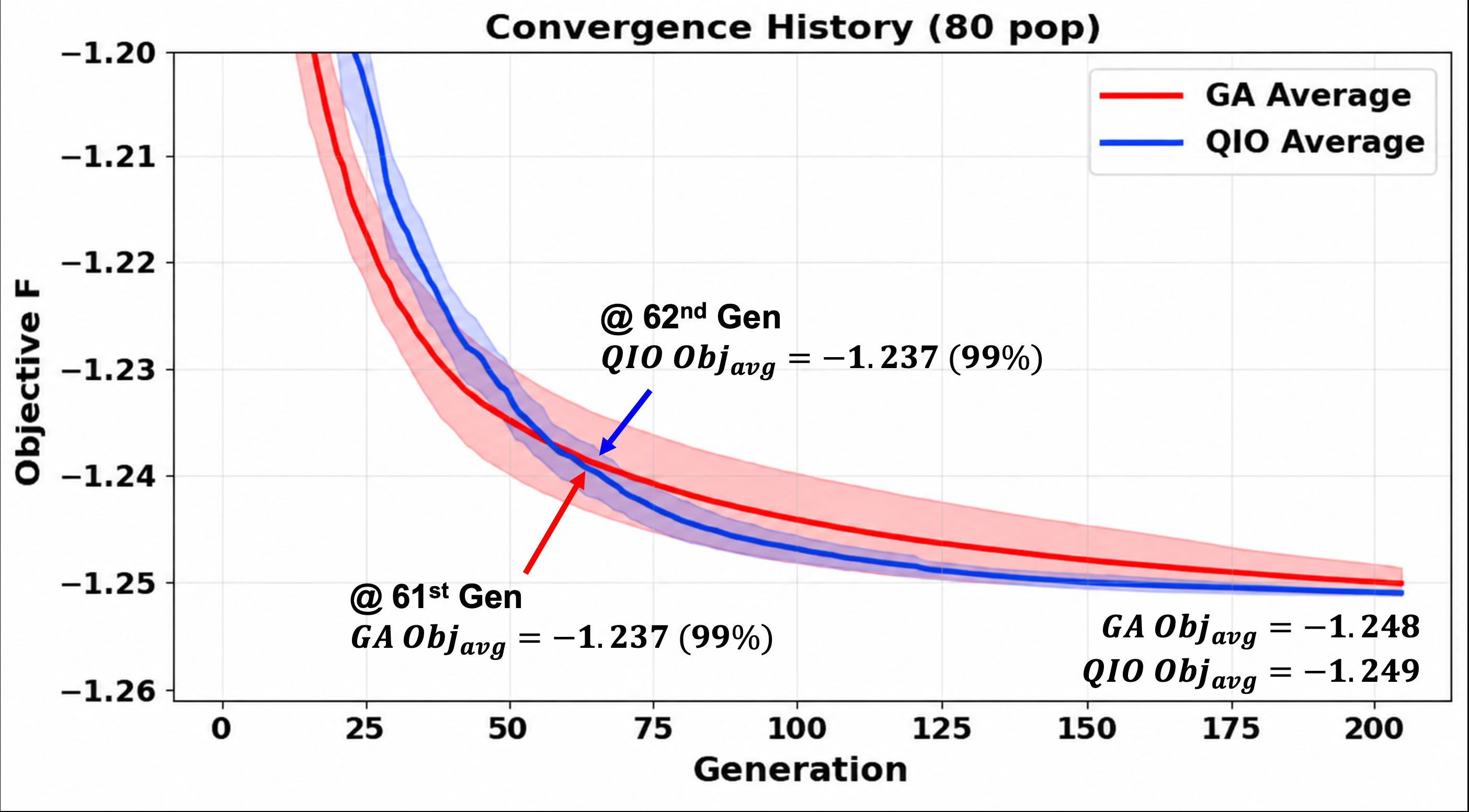}
    \caption{Convergence histories of GA and QIO for the large population case}
    \label{fig:airfoil_convergence_large}
\end{figure}

\begin{figure}[hbt]
    \centering
    \includegraphics[width=0.85\linewidth]{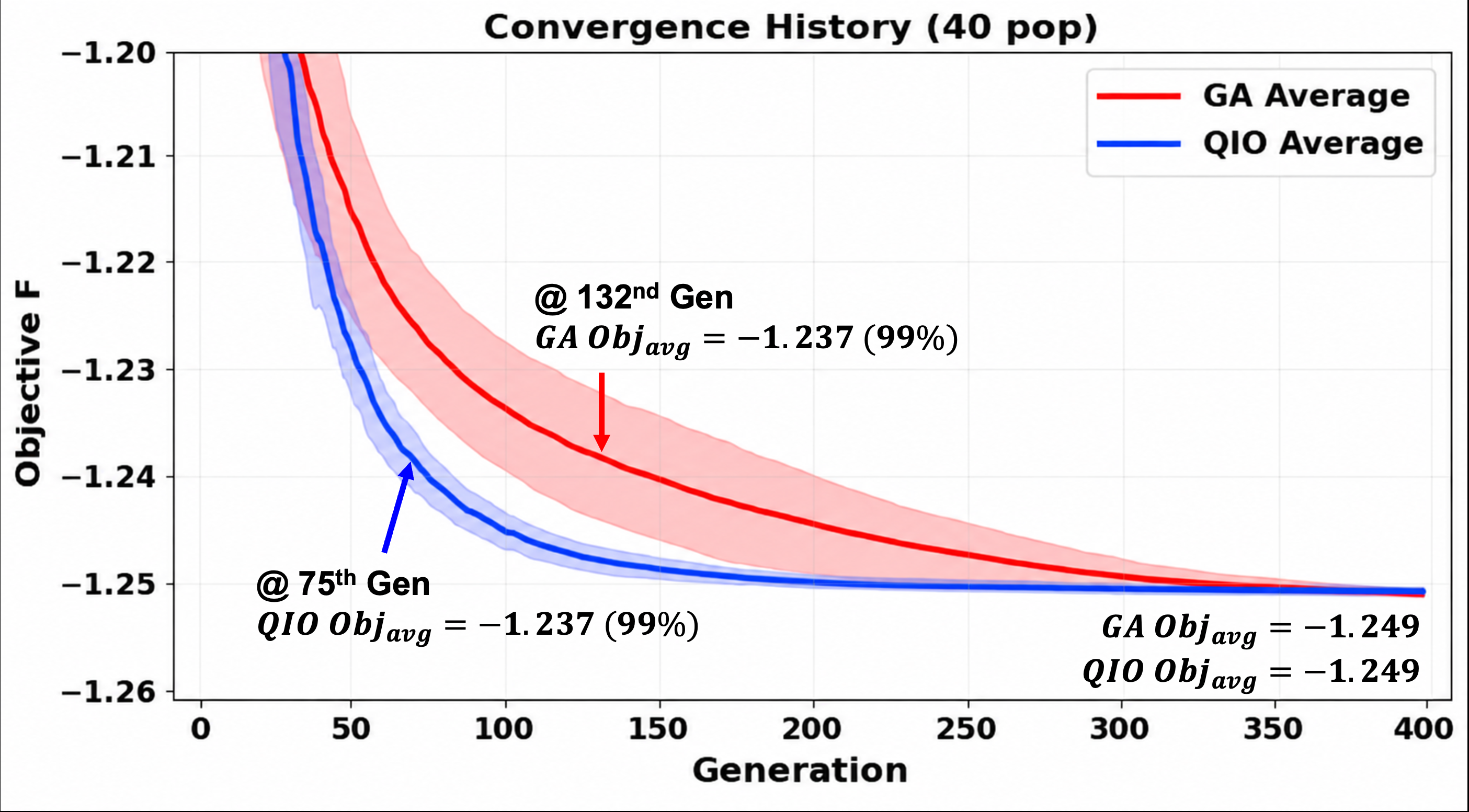}
    \caption{Convergence histories of GA and QIO for the moderate population case}
\label{fig:airfoil_convergence_moderate}
\end{figure}

\begin{figure}[hbt]
    \centering
    \includegraphics[width=0.85\linewidth]{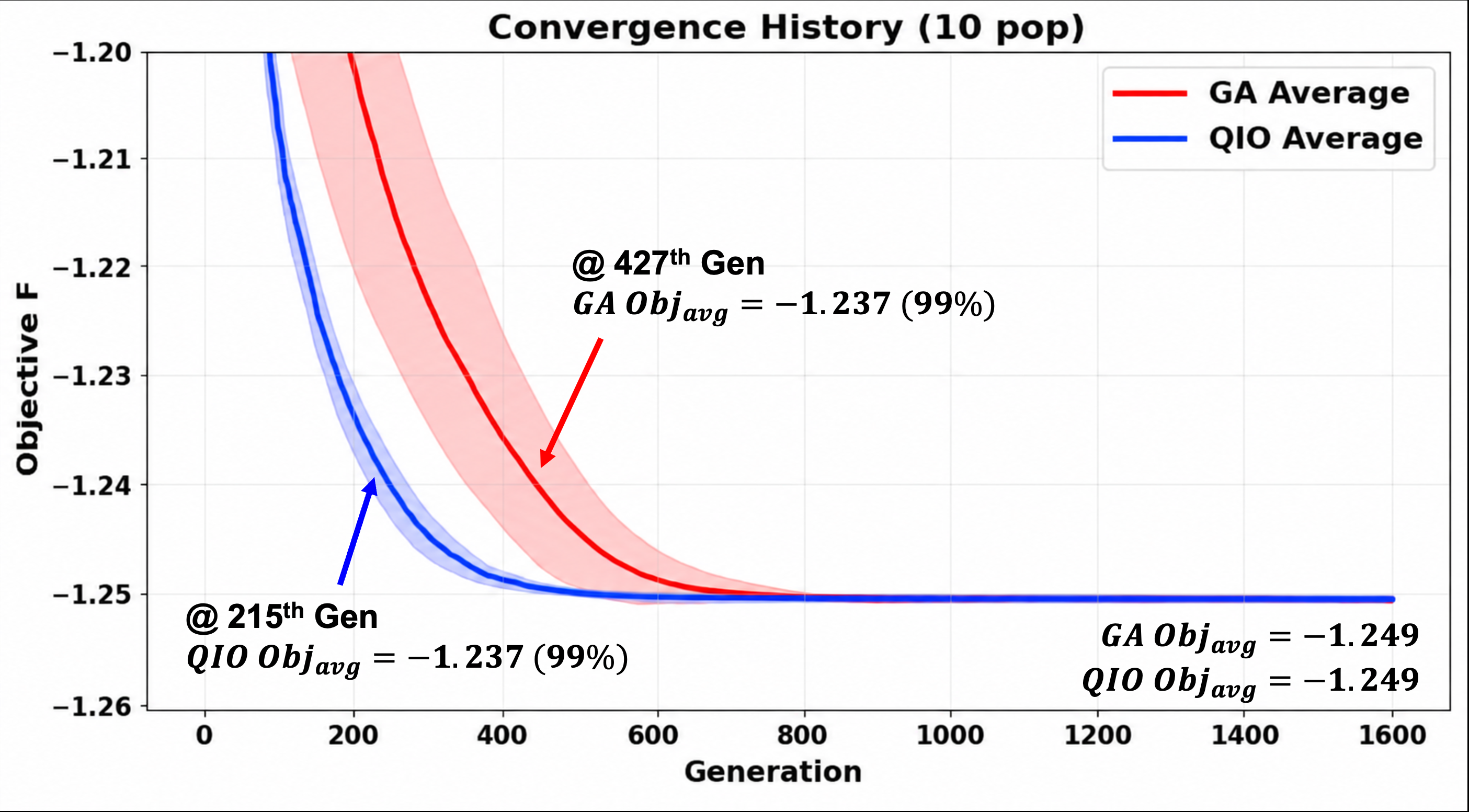}
    \caption{Convergence histories of GA and QIO for the small population case}
    \label{fig:airfoil_convergence_small}
\end{figure}

The airfoil optimization problem was first used to evaluate the convergence behavior of GA and QIO under different population-size settings. In all cases, the total number of function evaluations was fixed at 16,000, and each optimization was repeated using 15 different random seeds. The solid lines in the convergence plots represent the mean objective value over the independent runs, while the shaded regions indicate one standard deviation.

Figure \ref{fig:airfoil_convergence_large} shows the convergence histories for the large population case with 80 individuals and 200 generations. In this case, both GA and QIO converged to nearly identical final objective values. The number of generations required to reach 99\% of the best objective value was also similar: GA required 61 generations, while QIO required 62 generations. For this population size, the convergence behavior of GA and QIO was nearly identical, while QIO exhibited a smaller standard deviation.

As the population size decreased, the difference between the two optimizers became more pronounced. Figure \ref{fig:airfoil_convergence_moderate} shows the moderate-population case with 40 individuals and 400 generations. Although GA and QIO again reached comparable final objective values, QIO reached the near-optimal region faster. Specifically, QIO reached 99\% of the best objective value in 75 generations, whereas GA required 132 generations. Since both methods used the same population size in this case, this corresponds to 3,000 function evaluations for QIO and 5,280 function evaluations for GA. In addition, the QIO runs also remained more tightly clustered over most of the convergence history.

The advantage of QIO became more significant in the small-population case, as shown in Fig.~\ref{fig:airfoil_convergence_small}. With only 10 individuals, GA required 427 generations to reach 99\% of the best objective value, while QIO required only 215 generations. This corresponds to 4,270 function evaluations for GA and 2,150 function evaluations for QIO. Therefore, QIO reduced the required number of function evaluations by approximately 49.6\% compared with GA for this reduced-population setting. Also, the spread between independent runs was larger for GA, particularly before convergence, whereas QIO maintained a narrower standard deviation band throughout most of the optimization.

The evaluation efficiency of the two optimizers is summarized in Table~\ref{tab:airfoil_99obj}. Across all population-size cases, both GA and QIO converged to final objective values close to \(-1.249\).
Consequently, the reference objective values used to define the 99\% convergence criterion were nearly identical across the different optimizer and population-size cases. For the large-population case, QIO required a slightly larger number of evaluations than GA to reach this criterion. However, as the population size decreased, QIO required significantly fewer evaluations. In the moderate- and small-population cases, QIO reduced the required number of evaluations by 43.2\% and 49.6\%, respectively.

\begin{table}[H]
    \centering
    \caption{Number of generations and evaluations required to reach 99\% of the best airfoil objective value}
    \label{tab:airfoil_99obj}
    \begin{tabular}{lcccccc}
        \hline
        & \multicolumn{2}{c}{Large population} 
        & \multicolumn{2}{c}{Moderate population} 
        & \multicolumn{2}{c}{Small population} \\
        \cline{2-7}
        & GA & QIO & GA & QIO & GA & QIO \\
        \hline
        Generations for 99\% objective & 61 & 62 & 132 & 75 & 427 & 215 \\
        Function evaluations & 4880 & 4960 & 5280 & 3000 & 4270 & 2150 \\
        Relative change in evaluations & -- & +1.64\% & -- & -43.2\% & -- & -49.6\% \\
        \hline
    \end{tabular}
\end{table}

\begin{figure}[htbp]
    \centering

    \begin{subfigure}[t]{0.6\linewidth}
        \centering
        \includegraphics[width=\linewidth]
        {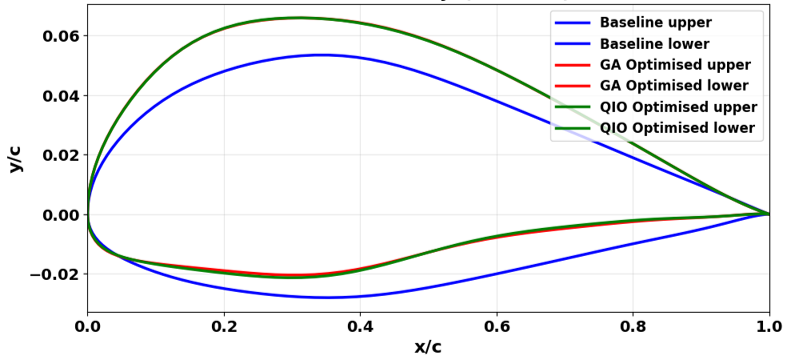}
        \caption{Large population case}
        \label{fig:airfoil_design_large}
    \end{subfigure}

    \vspace{0.4em}

    \begin{subfigure}[t]{0.6\linewidth}
        \centering
        \includegraphics[width=\linewidth]
        {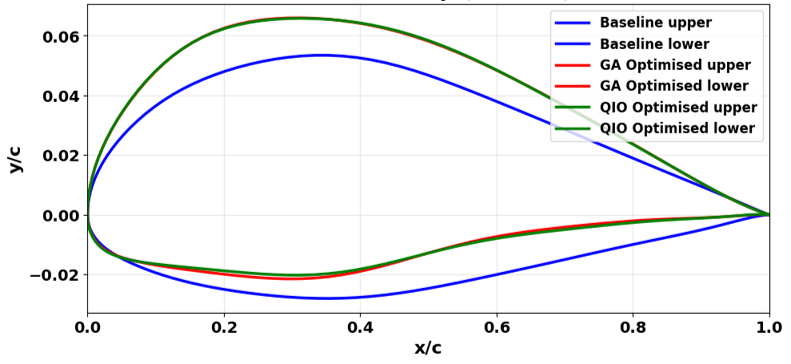}
        \caption{Moderate population case}
        \label{fig:airfoil_design_moderate}
    \end{subfigure}

    \vspace{0.4em}

    \begin{subfigure}[t]{0.6\linewidth}
        \centering
        \includegraphics[width=\linewidth]
        {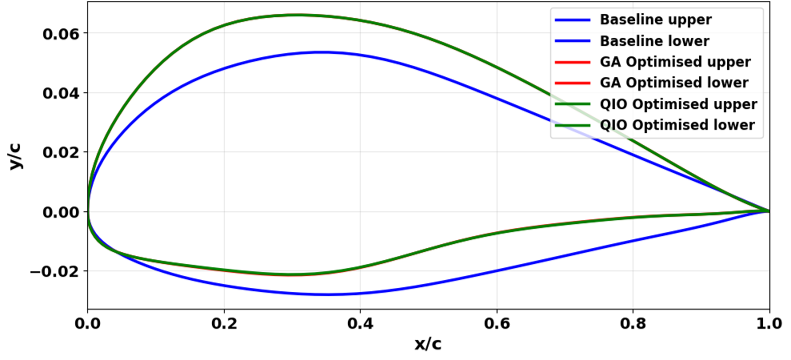}
        \caption{Small population case}
        \label{fig:airfoil_design_small}
    \end{subfigure}

    \caption{Comparison of the baseline airfoil and the optimized 
    airfoil geometries obtained using GA and QIO under different 
    population-size settings}
    \label{fig:airfoil_design_results}
\end{figure}

The optimized airfoil geometries are compared in Fig.~\ref{fig:airfoil_design_results}. Overall, the final airfoil shapes obtained by GA and QIO are similar across all population sizes. Both methods therefore reached comparable final designs under the fixed evaluation budget. However, the convergence histories show that QIO reaches these high-quality designs using fewer evaluations when the population size is limited.

Taken together, the convergence and geometry results in Figs. \ref{fig:airfoil_convergence_large}--\ref{fig:airfoil_design_results} show that the primary advantage of QIO lies in its efficient use of a limited population. When a sufficiently large population is available, the crossover and mutation operations of GA appear to provide adequate diversity, leaving little difference between the two optimizers. As the population size decreases, however, the GA operators act on a smaller set of deterministic candidate solutions, which can limit the amount of new information introduced in each generation and slow progress toward the optimal design region.

In QIO, each chromosome represents a probability distribution rather than a single deterministic solution. Updating this distribution allows information associated with alternative states to be retained while the sampling probability is gradually concentrated around promising solutions. This mechanism is consistent with the observed population-size dependence, where its effect is limited when many individuals are available but becomes increasingly beneficial as the population is reduced.

To further assess the quality of the final airfoil designs, the final designs obtained by GA and QIO were used to initialize a gradient-based optimization using sequential least squares programming (SLSQP). The refinement retained the original objective function, design-variable bounds, lift constraint, and thickness-normalization procedure. As summarized in Table~\ref{tab:airfoil_slsqp_refinement}, the mean improvement over the 15 random seeds was below $0.2\%$ for both optimizers in every population-size case. Although global optimality cannot be established, the final solutions from both GA and QIO were close to those from SLSQP.

\begin{table}[htbp]
    \centering
    \caption{Mean refined objective and objective improvement obtained by SLSQP}
    \label{tab:airfoil_slsqp_refinement}
    \begin{tabular}{lcccc}
        \toprule
        Population size
        & Optimizer
        & Refined objective
        & Mean Improvement (\%)
        & Maximum Improvement (\%) \\
        \midrule
        Large (80)
        & GA
        & $-1.2492 \pm 3.75\times10^{-7}$
        & 0.04592
        & 0.28116 \\
        
        & QIO
        & $-1.2492 \pm 1.71\times10^{-6}$
        & 0.12839
        & 0.21944 \\
        \midrule
        
        Moderate (40)
        & GA
        & $-1.2492 \pm 2.68\times10^{-6}$
        & 0.02963
        & 0.17191 \\
        
        & QIO
        & $-1.2492 \pm 5.16\times10^{-7}$
        & 0.01999
        & 0.03708 \\
        \midrule
        
        Small (10)
        & GA
        & $-1.2492 \pm 9.11\times10^{-8}$
        & $2.47\times10^{-5}$
        & $4.98\times10^{-5}$ \\
        
        & QIO
        & $-1.2492 \pm 7.68\times10^{-7}$
        & 0.01244
        & 0.04377 \\
        \bottomrule
    \end{tabular}
\end{table}

\subsection{Rotor Blade Optimization Results}
\begin{figure}[H]
    \centering
    \includegraphics[width=0.85\linewidth]{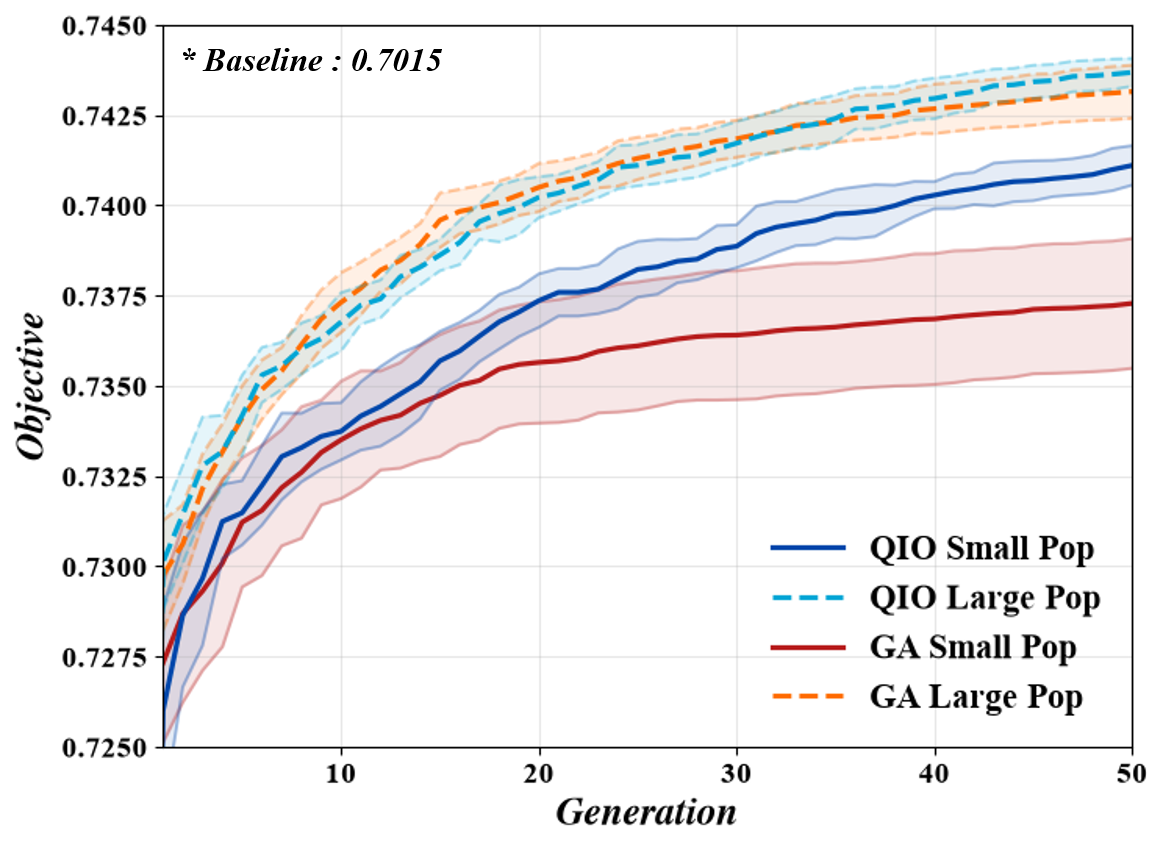}
    \caption{Comparison of the convergence histories of GA and QIO}
    \label{fig:rbo_convergence}
\end{figure}

The rotor blade optimization problem was used to evaluate the performance of GA and QIO in a higher-dimensional design space. Figure~\ref{fig:rbo_convergence} compares the mean convergence histories of GA and QIO for the small and large population cases. Each curve represents the average of ten independent runs, and the shaded region denotes one standard deviation across the runs.

The figure of merit results at the last generation are summarized in Table \ref{tab:rotor_fm_results}. For the large-population cases, GA and QIO exhibited similar convergence behavior and reached comparable final objective values. QIO obtained an average final FM of 0.7437, while GA obtained an average FM of 0.7432. These values correspond to improvements of 6.02\% and 5.94\%, respectively, relative to the baseline blade.

The effect of the optimizer became more apparent when the population was reduced from 256 to 30. The small-population QIO case reached an average final FM of 0.7411, corresponding to a 5.65\% improvement over the baseline. In comparison, the small-population GA case reached an average FM of 0.7373, corresponding to a 5.10\% improvement. The QIO result therefore remained closer to the large-population solutions despite using fewer populations per generation.

QIO also exhibited lower seed-to-seed variability. The standard deviation of the final FM was 0.0006 for small-population QIO, compared with 0.0019 for small-population GA. The difference in standard deviation between GA and QIO was smaller in the large-population cases. A larger population provides broader coverage of the design space and maintains greater population diversity, thereby reducing the sensitivity of both optimizers to the initial random seed.

\begin{table}[H]
    \centering
    \caption{Average figure of merit obtained from rotor blade optimization}
    \label{tab:rotor_fm_results}
    \begin{tabular}{lccc}
        \hline
        Case & Optimizer & Average FM & Improvement over baseline \\
        \hline
        Baseline & -- & 0.7015 & -- \\
        Small population & QIO  & 0.7411 & +5.65\% \\
        Small population & GA & 0.7373 & +5.10\% \\
        Large population & QIO  & 0.7437 & +6.02\% \\
        Large population & GA & 0.7432 & +5.94\% \\
        \hline
    \end{tabular}
\end{table}

\begin{figure}[htbp]
    \centering

    \begin{subfigure}[b]{0.48\linewidth}
        \centering
        \includegraphics[width=\linewidth]
        {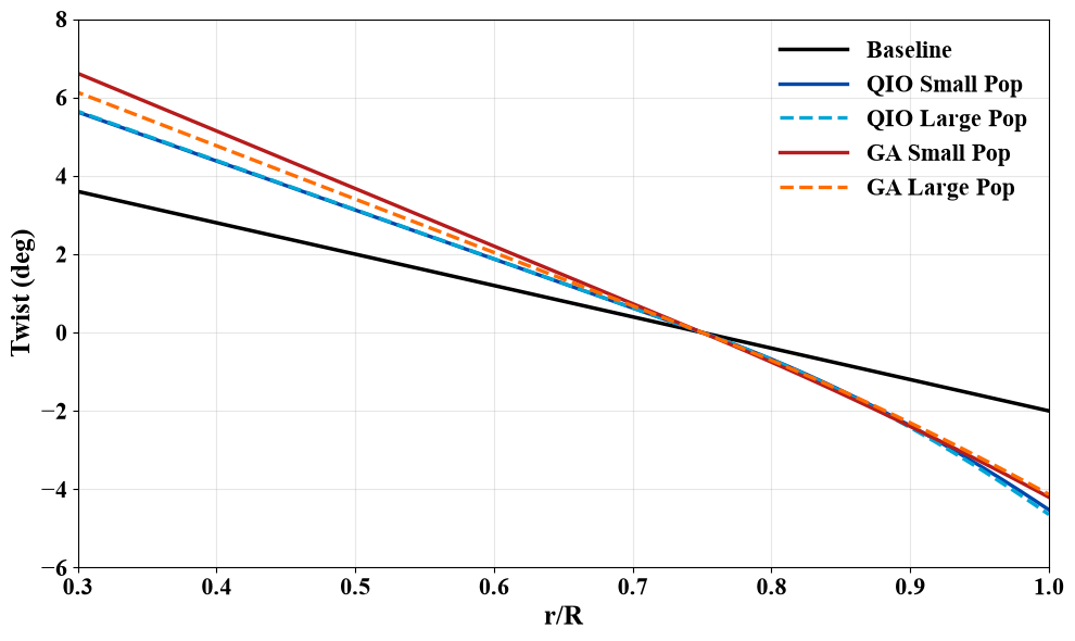}
        \caption{Twist distribution}
        \label{fig:twist_distribution}
    \end{subfigure}
    \hfill
    \begin{subfigure}[b]{0.48\linewidth}
        \centering
        \includegraphics[width=\linewidth]
        {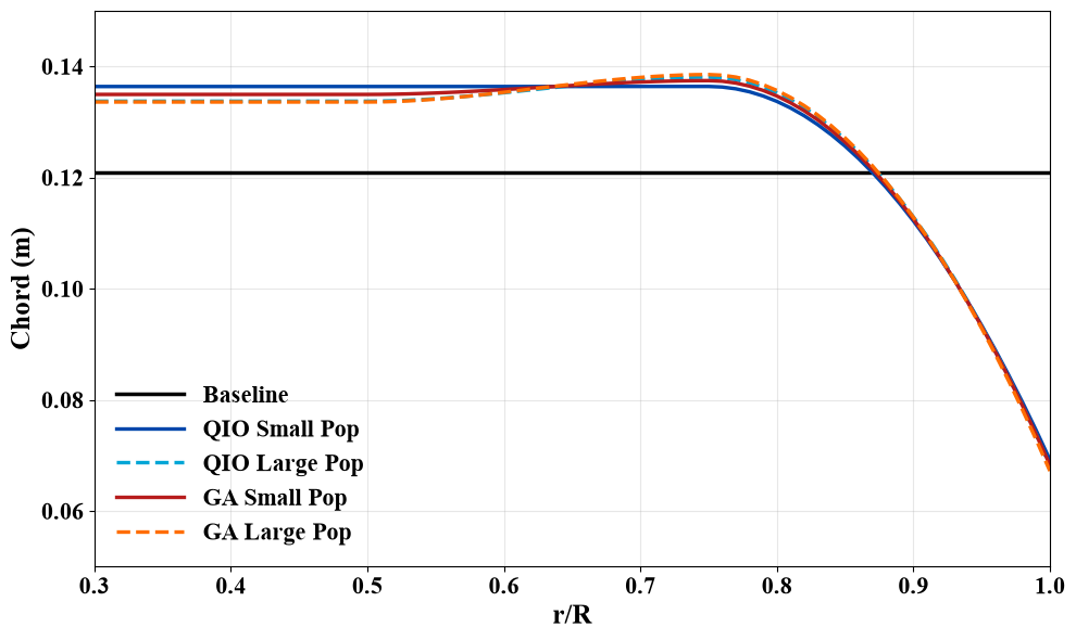}
        \caption{Chord distribution}
        \label{fig:chord_distribution}
    \end{subfigure}

    \caption{Comparison of optimized blade planform distributions}
    \label{fig:rbo_planform}
\end{figure}

Figure~\ref{fig:rbo_planform} compares the twist and chord distributions of the best-performing blade from each fresh-start optimization case. All optimized blades exhibit substantially greater twist variation than the baseline blade. The optimized twist distributions are similar near $r/R=0.75$, but small differences develop toward the root and tip. The small-population GA design has the highest root twist and the least-negative tip twist, whereas the QIO solutions remain closer to the corresponding large-population trends.

Likewise, the optimized chord distributions are different from the rectangular baseline planform but are similar among the four optimization cases. The chord remains nearly constant over the inboard portion of the blade, increases slightly near $r/R=0.7$, and then decreases rapidly toward the tip. The agreement among the optimized chord distributions indicates that both optimizers consistently identify a similar planform trend.

\begin{figure}[htbp]
    \centering

    \begin{subfigure}[b]{0.48\linewidth}
        \centering
        \includegraphics[width=\linewidth]
        {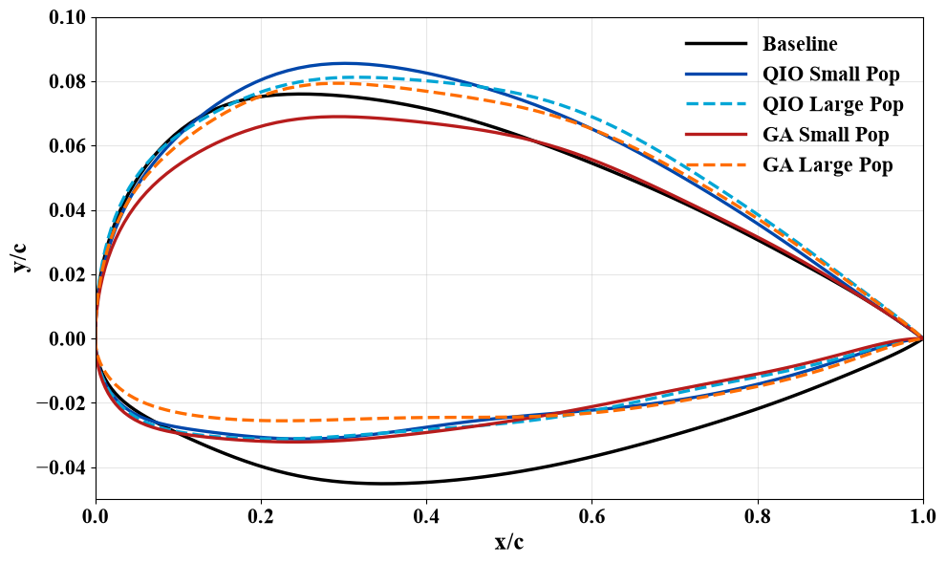}
        \caption{Inboard airfoil}
        \label{fig:rbo_airfoil_geometry_inboard}
    \end{subfigure}
    \hfill
    \begin{subfigure}[b]{0.48\linewidth}
        \centering
        \includegraphics[width=\linewidth]
        {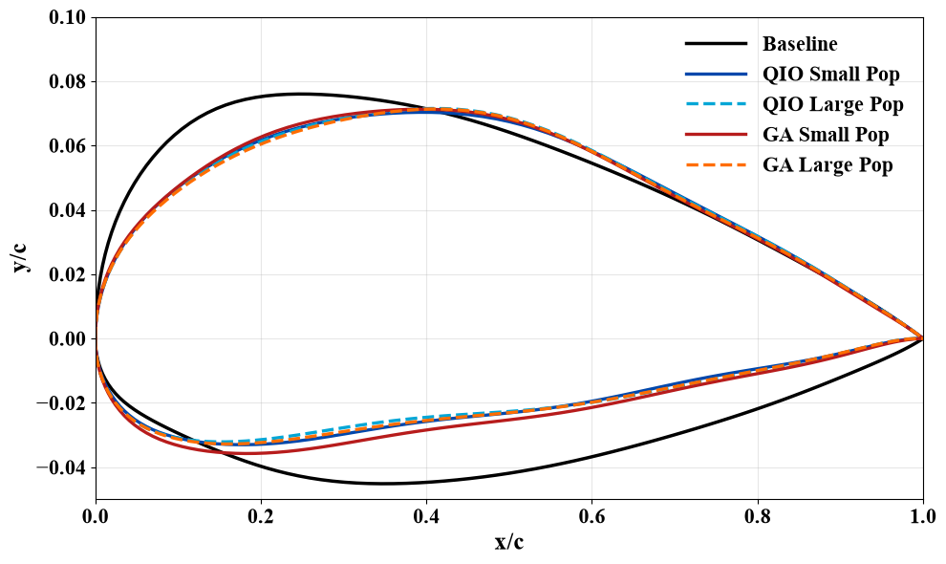}
        \caption{Outboard airfoil}
        \label{fig:rbo_airfoil_geometry_outboard}
    \end{subfigure}

    \caption{Comparison of optimized inboard and outboard airfoil geometries}
    \label{fig:rbo_airfoil}
\end{figure}

Figure~\ref{fig:rbo_airfoil} compares the optimized inboard and outboard airfoil sections. The inboard sections exhibit greater variation than the outboard sections where small-population GA produces a thinner inboard section than the other solutions. In contrast, the two QIO geometries remain similar to each other despite the difference in population size. For the outboard sections, all four optimized airfoils have similar geometries with reduced thickness and with more camber.

\begin{figure}[H]
    \centering
    \includegraphics[width=0.6\linewidth]{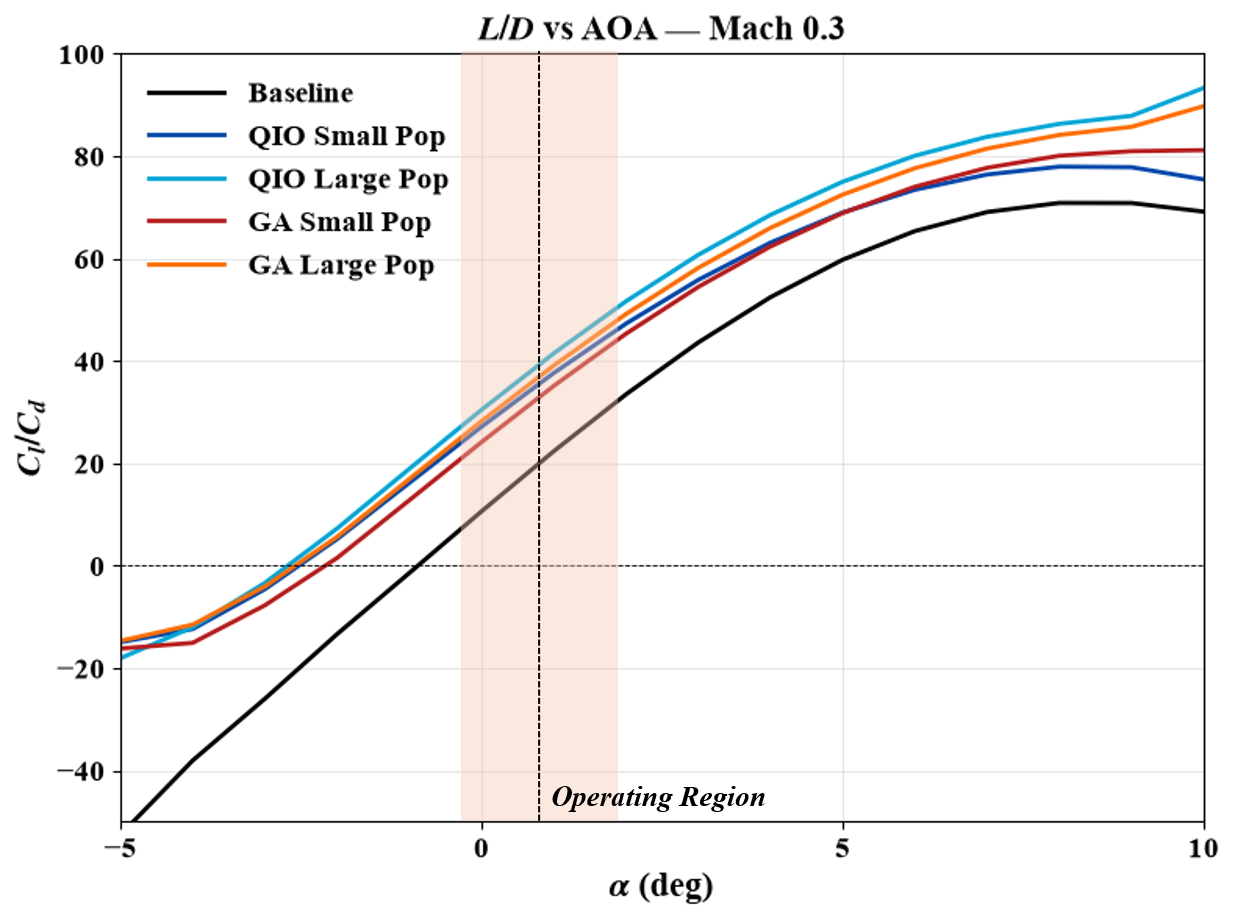}
    \caption{Aerodynamic performance comparison of optimized inboard airfoils}
    \label{fig:rbo_airfoil_inboard}
\end{figure}

To further examine the aerodynamic implications of the optimized airfoil shapes, the sectional airfoil performance was compared for both the inboard and outboard airfoils. Figure~ \ref{fig:rbo_airfoil_inboard} shows the performance comparison for the inboard airfoil. The inboard region experiences relatively low dynamic pressure, but the combination of collective pitch and blade twist can result in higher local angles of attack. Therefore, maintaining good aerodynamic performance over a wide range of angles of attack is important. At Mach 0.3, all optimized airfoils provide a higher lift-to-drag ratio than the baseline airfoil throughout the highlighted operating region. The large-population QIO design generally provides the highest sectional efficiency, while the other optimized designs show similar trends near the design angle of attack.

\begin{figure}[H]
    \centering
    \includegraphics[width=0.6\linewidth]{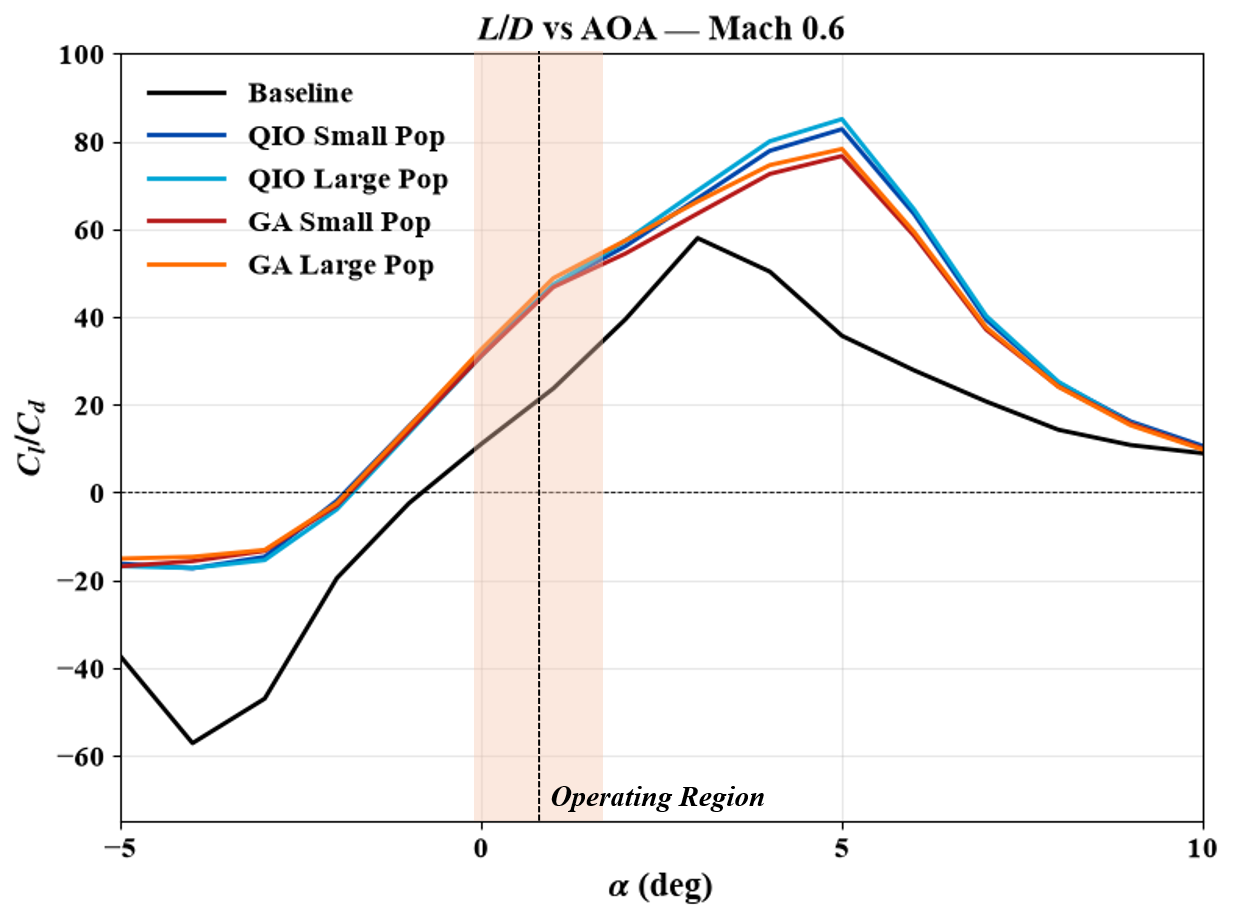}
    \caption{Aerodynamic performance comparison of optimized outboard airfoils}
    \label{fig:rbo_airfoil_outboard}
\end{figure}

In contrast, the outboard region experiences higher dynamic pressure and operates near Mach 0.6. Under these conditions, reducing aerodynamic drag becomes particularly important for rotor efficiency. As shown in Fig.~\ref{fig:rbo_airfoil_outboard}, all optimized outboard airfoils achieve higher lift-to-drag ratios than the baseline within the operating region. The performance differences among the four optimized designs are relatively small, which is consistent with their similar airfoil geometries. The reduced thickness of the optimized sections is also consistent with the need to limit drag penalties at higher Mach numbers.

Overall, the geometry and sectional-performance comparisons support the convergence results. The large-population GA and QIO cases identify similar high-performance blade designs. When the population is reduced, QIO retains the principal planform and airfoil characteristics of the large-population solutions and achieves a final FM closer to the large-population reference. The small-population GA reaches the same general design region but exhibits greater seed-to-seed variability and a lower average final objective.

\subsection{Comparison with Neural-Network-Based Blade Design}

\begin{figure}[H]
    \centering
    \includegraphics[width=0.82\linewidth]{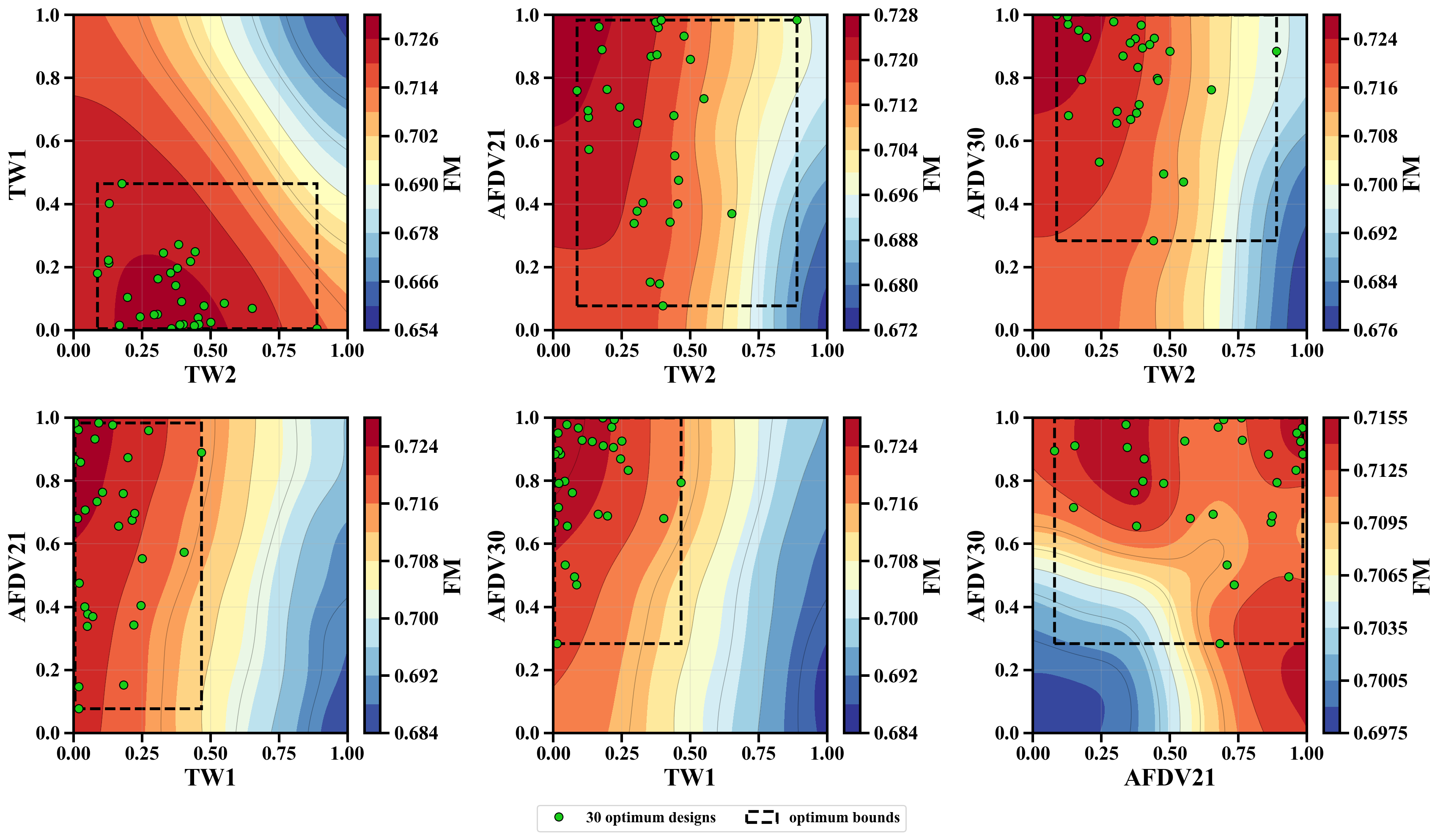}
    \caption{Neural-network-identified promising regions and selected points}
    \label{fig:hybrid_promising_region}
\end{figure}

The surrogate-informed hybrid approach was evaluated by using the neural-network-identified designs as the initial population of the small-population GA. Figure~\ref{fig:hybrid_promising_region} shows the neural-network-identified promising region projected onto pairs of influential variables. The green markers denote the 30 high-ranked feasible designs used to identify the high-FM neighborhood, and the dashed rectangles show the corresponding projected bounds. The selected points occupy a portion of the design space where FM is predicted as high.

\begin{figure}[H]
    \centering
    \includegraphics[width=0.7\linewidth]{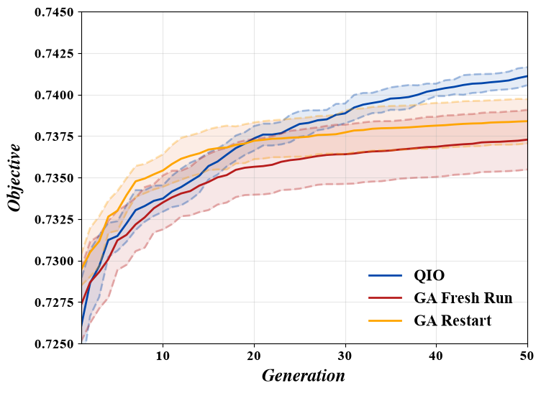}
    \caption{Comparison of fresh-start QIO, fresh-start GA, and neural-network-initialized GA}
    \label{fig:rbo_hybrid_convergence}
\end{figure}

Figure~\ref{fig:rbo_hybrid_convergence} compares the corresponding convergence histories. The NN-informed initialization increased the initial GA objective from 0.7270 to 0.7290 on average. At Generation 5, the hybrid GA reached an average FM of 0.7333, compared with 0.7313 for the fresh-start GA. At Generation 50, the hybrid GA reached an average FM of 0.7380, compared with 0.7374 for fresh-start GA. Thus, the hybrid initialization retained a final improvement of approximately 0.0006 in FM over the conventional GA, but the largest improvement occurred during the early generations.

Fresh-start QIO initially had a lower objective than the hybrid GA. However, the mean QIO convergence curve intersected the hybrid GA curve at approximately Generation 17 and subsequently achieved a higher final objective value. At Generation 50, fresh-start QIO obtained an average FM of 0.7409, exceeding the hybrid GA result by approximately 0.0029.

\begin{figure}[H]
    \centering
    \includegraphics[width=0.7\linewidth]{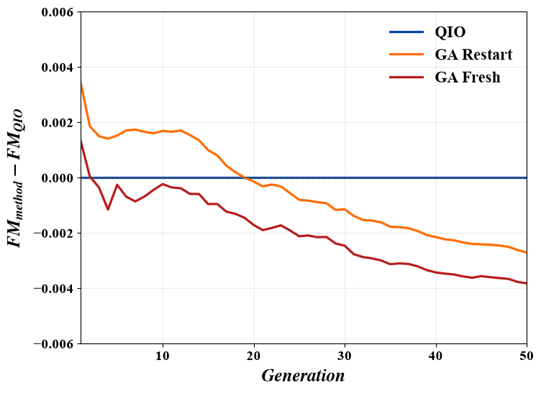}
    \caption{Difference in mean figure of merit to the QIO small population case}
    \label{fig:rbo_hybrid_delta}
\end{figure}

Figure~\ref{fig:rbo_hybrid_delta} shows the generation-dependent FM difference relative to fresh-start QIO. Positive values indicate that a method has a higher average FM than QIO at the corresponding generation. The hybrid GA provides an initial advantage of approximately 0.002--0.003 in FM, whereas the fresh-start GA remains closer to or below QIO during most of the optimization. After approximately 17 generations, both GA cases fall below the QIO curve.

These results show that surrogate-informed initialization successfully accelerates the initial global search stage of the GA. However, it does not completely compensate for the slower subsequent convergence of the reduced-population GA. Fresh-start QIO requires no surrogate-training stage and ultimately achieves a higher objective than the hybrid GA. Therefore, the hybrid method is most beneficial when GA is used with a limited evaluation budget, whereas fresh-start QIO yields the better final solution under the current 50-generation budget.

\section{Conclusion}
This study investigated the application of quantum-inspired optimization to aerodynamic design problems relevant to rotor blade optimization. The performance of QIO was compared with that of a conventional genetic algorithm using two design problems: an airfoil optimization problem evaluated with a neural-network surrogate model and a rotor blade optimization problem evaluated with UMARC-II.

For the airfoil optimization problem, GA and QIO achieved comparable final airfoil designs across all population-size settings. In the large-population case, both methods showed nearly identical convergence behavior. However, as the population size was reduced, QIO reached near-optimal solutions with fewer function evaluations. In the moderate-population case, QIO reduced the required number of evaluations by 43.2\% relative to GA, while in the small-population case, the reduction reached 49.6\%.

For the rotor blade optimization problem, both GA and QIO improved the hover figure of merit relative to the baseline HART-II blade. In the large-population case, the two optimizers achieved similar mean final FM values of 0.7432 for GA and 0.7437 for QIO. When the population was reduced from 256 to 30, the mean final FM decreased to 0.7373 for GA and 0.7411 for QIO. The corresponding standard deviations were 0.0019 and 0.0006, respectively, indicating that the reduced-population QIO result was less sensitive to the initial random seed. The selected small-population QIO blade also retained the principal planform and airfoil characteristics of the large-population solutions more closely than the selected small-population GA blade.

The surrogate-informed initialization improved the early convergence of the reduced-population GA. The mean FM at first generation increased from 0.7270 for fresh GA to 0.7290 for hybrid GA, and the final mean FM increased from 0.7374 to 0.7380. Fresh-start QIO exceeded the hybrid GA at Generation 20 and reached a higher final mean FM of 0.7409 without the additional 250 UMARC-II evaluations used to construct the neural-network surrogate.

The results suggest that the primary advantage of QIO is not a substantially different final design when a sufficiently large population is available. Rather, QIO provides greater evaluation efficiency and lower seed-to-seed variability as the population size is reduced. This behavior is particularly relevant to rotor blade design problems in which each function evaluation may require a higher-fidelity aerodynamic or coupled CFD--CSD analysis. The surrogate-informed approach remains useful when an existing model can be reused or when only a small number of refinement generations is affordable, whereas fresh-start QIO provides the higher final performance under the present optimization budget.

\section{Future Work}

Future work will extend the present formulation to multi-objective optimization that considers both hover and forward-flight performance. Additional aerodynamic and structural design variables will also be incorporated to enable a more comprehensive rotor-blade design. Finally, the evaluation efficiency observed for QIO motivates its application to higher-fidelity optimization using free-wake and CFD-based analyses, for which reducing the number of required function evaluations becomes increasingly important.

\bibliography{reference}

\section{Appendix}
\subsection{Effect of Neural-Network Initialization on QIO}
\begin{figure}[h]
    \centering
    \includegraphics[width=0.65\linewidth]{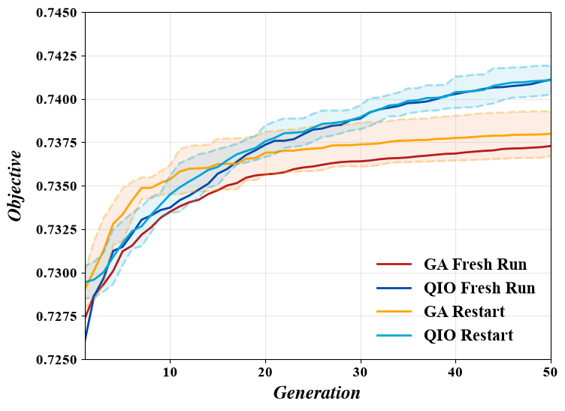}
    \caption{Comparison of fresh-start and neural-network-initialized restart cases for GA and QIO}
    \label{fig:rbo_restart_convergence_all}
\end{figure}

{The fresh and restart QIO cases are compared in this section. Figure~\ref{fig:rbo_restart_convergence_all} compares the convergence histories of fresh-start and neural-network-initialized restart cases for GA and QIO. The QIO restart used the same initial population as the surrogate-informed hybrid GA approach. While this initialization improved GA's early performance, the QIO restart and fresh-start histories remained nearly identical after the first few generations. Across the cases examined, fresh-start QIO required approximately five generations on average to reach the objective value provided by the seeded population, compared with 10.8 generations for fresh-start GA.}

{In the case of GA, GA directly evolves explicitly encoded design vectors, so a seeded population places the initial chromosomes within the neural-network-predicted promising region. This information can remain influential through selection and recombination during the early generations. However, QIO represents each design variable as a probability amplitude and generates candidate solutions by sampling from the resulting probabilistic representation. The seeded designs, therefore, do not remain fixed chromosomes throughout the search. Their information influences the initial probability representation, which is subsequently modified as QIO updates its search distribution. In the present problem, fresh-start QIO recovered a comparable objective level within only a few generations, leaving little opportunity for the restart population to provide a lasting advantage.}

\subsection{Extended Convergence of the Small-Population Cases}

\begin{figure}[H]
\centering
\includegraphics[width=0.72\linewidth]{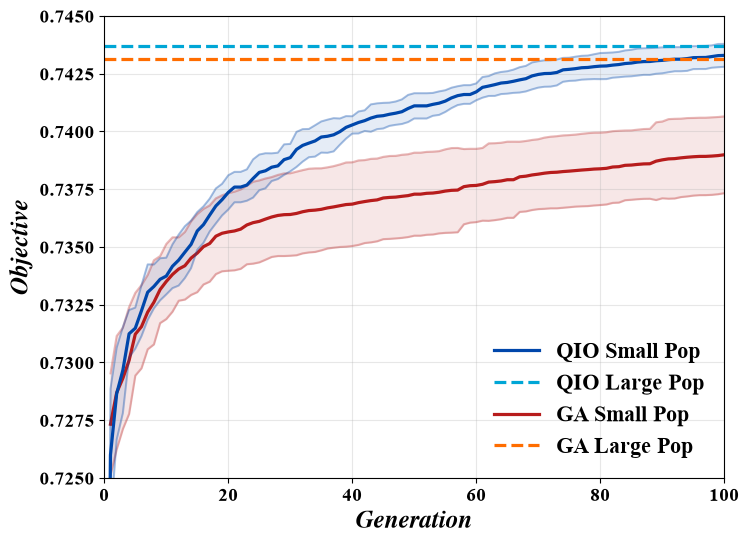}
\caption{Extended convergence of the small-population GA and QIO cases}
\label{fig:rbo_extended_convergence}
\end{figure}

To examine whether the performance loss associated with population reduction could be recovered with a larger optimization budget, the small-population GA and QIO cases were continued for an additional 50 generations, resulting in a total of 100 generations. Figure~\ref{fig:rbo_extended_convergence} compares the extended mean convergence histories with the final mean objectives of the corresponding large population cases, shown as horizontal dashed lines. The solid curves and shaded bands represent the mean and one standard deviation over the 10 random seeds, respectively.

After 100 generations, the small-population QIO case reached a mean FM of 0.7433, approaching the large-population QIO value of 0.7437 obtained after 50 generations. The additional generations therefore allowed QIO to recover nearly all of the objective gap caused by the reduced population size. In contrast, the small-population GA case reached a mean FM of 0.7390 and remained below the large-population GA value of 0.7432. Thus, increasing the number of generations did not compensate for the loss in convergence performance caused by population reduction in GA. 

\end{document}